\documentclass[graphics,floatfix, nofootinbib,tightenlines,nobibnotes, 12pt]{revtex4-2}

\usepackage{hyperref}
\usepackage{graphicx}
\usepackage[english]{babel}
\usepackage[utf8]{inputenc}
\usepackage{amsmath,amssymb}
\usepackage{amsfonts}
\usepackage{multirow}
\usepackage{pstricks}
\usepackage{float}
\usepackage{subfigure}
\usepackage{color}
\usepackage{epsfig}
\usepackage{slashed}

\begin{document}

\title{Production of Excited Doubly Heavy Baryons at the Super-$Z$ Factory}

\author{Juan-Juan Niu$^{1,2}$}
\email{niujj@gxnu.edu.cn}
\author{Jing-Bo Li$^{1}$}
\author{Huan-Yu Bi$^{3,4}$}
\email{bihy@pku.edu.cn}
\author{Hong-Hao Ma$^{1,2}$}
\email{mahonghao@pku.edu.cn, corresponding author}

\address{$^{1}$ Department of Physics, Guangxi Normal University, Guilin 541004, China}
\address{$^{2}$ Guangxi Key Laboratory of Nuclear Physics and Technology, Guangxi Normal University, Guilin 541004, China}
\address{$^{3}$ Center for High Energy Physics, Peking University, Beijing 100871, China}
\address{$^{4}$ School of Physics, Peking University, Beijing 100871, China}

\date{\today}

\begin{abstract}

In the framework of nonrelativistic QCD, the excited doubly heavy baryons are thoroughly studied  via the channel $e^{+} e^{-}\rightarrow \langle QQ^{\prime}\rangle[n] \rightarrow \Xi_{QQ^{\prime}} +\bar{Q^{\prime}} +\bar{Q}$, which takes place at the collision energy $Z$-pole. $Q^{(\prime)}$ represents $b$ or $c$ quark for the production of $\Xi_{cc}$, $\Xi_{bc}$, and $\Xi_{bb}$, respectively. All of the intermediate diquark states $\langle QQ'\rangle[n]$ in $P$-wave, $\langle cc\rangle[^{1}P_{1}]_{\mathbf{\bar 3}}$, $\langle cc\rangle[^{3}P_{J}]_{\mathbf{6}}$, 
$\langle bc\rangle[^{1}P_{1}]_{\mathbf{\bar 3}/ \mathbf{6}}$, $\langle bc\rangle[^{3}P_{J}]_{\mathbf{\bar 3}/ \mathbf{6}}$, 
$\langle bb \rangle[^{1}P_{1}]_{\mathbf{\bar 3}}$, and $\langle bb\rangle[^{3}P_{J}]_{\mathbf{6}}$ with $J=0$, 1, or 2, are taken into account. 
The cross sections and differential distributions, including the transverse momentum, rapidity, angular, and invariant mass, are discussed for the excited baryons production. We find that the contributions of $\langle cc \rangle$, $\langle bc \rangle$, and $\langle bb \rangle$ in $P$-wave are found to be 3.97$\%$, 5.08$\%$, and 5.89$\%$, respectively, compared to $S$-wave. Supposing that all excited states can decay into the ground state 100\%, the total events $N_{\Xi_{cc}}=8.48 \times10^{4-6}$, $N_{\Xi_{bc}}=2.26\times10^{5-7}$, and $N_{\Xi_{bb}}=4.12 \times10^{3-5}$ would be produced at the Super-$Z$ Factory with a high luminosity up to ${\cal L} \simeq 10^{34-36}{\rm cm}^{-2} {\rm s}^{-1}$.

\end{abstract}

\maketitle

\section{Introduction}

The quark model, which was introduced by Gell-Mann and Zweig in the 1960s \cite{GellMann:1964nj,Zweig:1981pd,Zweig:1964jf}, theoretically predicts the existence of doubly heavy baryons composed of two heavy quarks ($b$ or $c$) and a light quark ($u$, $d$ or $s$). The study about the production and decay of doubly heavy baryons can deepen our understanding of strong interactions. In 2017, the LHCb collaboration first experimentally observed the signal of the doubly charmed baryon $\Xi^{++}_{cc}$ \cite{Aaij:2017ueg}, providing direct and reliable evidence for the existence of doubly heavy baryons. Previous studies have mainly focused on nonrelativistic QCD (NRQCD)~\cite{Bodwin:1994jh,Falk:1993gb} and heavy quark effective field theories~\cite{Caswell:1985ui}, which have predicted some of the fundamental properties of doubly heavy baryons such as their mass, lifetime, production, and decay. 

NRQCD is an effective nonrelativistic theory that describes the interactions of quarks and antiquarks in hadronic systems, especially mesons and baryons containing heavy quarks $c$ or $b$. 
In hadronic system with nonrelativistic boundaries, the relative velocity $v$ between quarks is much smaller than the speed of light, enabling nonrelativistic methods to be applied. 
Within the framework of NRQCD, the production of doubly heavy baryons can be factorized into two parts, the perturbative region and non-perturbative region. In the perturbative region, the production mainly consists of a diquark state $\langle QQ^{\prime}\rangle[n]+X$, where $[n]$ is the spin and color quantum number of the diquark $\langle QQ^{\prime}\rangle$. Usually, direct production mechanisms such as strong production~\cite{Baranov:1995rc,Ma:2003zk,Chang:2006eu,Chang:2006xp} dominate the diquark production, supported by additional direct production mechanisms such as photon production~\cite{Baranov:1995rc,Chen:2014frw,Huan-Yu:2017emk,Li:2007vy} and electron-positron production~\cite{Jiang:2012jt,Jiang:2013ej,Zheng:2015ixa}, as well as indirect production mechanisms \cite{Liao:2018nab,Zhang:2022jst,Niu:2018ycb,Niu:2019xuq,Ma:2022ger,Ma:2022cgt}. In the non-perturbative region, the hadronization of the diquark to the doubly heavy baryon is primarily involved. Some important advances have been made in theoretical research based on lattice QCD, potential model, QCD sum rules, and so on \cite{Kiselev:2000jc,Kiselev:2002iy,Bodwin:1996tg,Kiselev:1999sc,Bagan:1994dy,Ali:2018ifm,Ali:2018xfq,Qin:2020zlg,Wang:2018lhz}. Due to the decoupling of $SU_C(3)$ color group $3\bigotimes3=\mathbf{\bar{3}}\bigoplus \mathbf{6}$ \cite{Zweig:1981pd,Zweig:1964jf}, the color quantum number of the diquark state can be color-antitriplet $\mathbf{\bar{3}}$ and -sextuplet $\mathbf{6}$.

The research on the production mechanisms of doubly heavy baryons are helpful to the study of particle properties and interactions. Among these production mechanisms, electron-positron production is a particularly efficient way because of the relatively clean background, high resolution and detection ability of the $e^{+} e^{-}$ collider. The future Super-$Z$ factory~\cite{Chang:2010SCPMA,Erler:2000jg} is a large experimental facility based on the $e^{+} e^{-}$ collider, CEPC or FCC-ee \cite{CEPCStudyGroup:2018ghi,FCC:2018evy}, whose collision energy is at the resonance $Z^0$ peak with a design luminosity up to ${\cal L} \simeq 10^{34-36}~{\rm cm}^{-2} {\rm s}^{-1}$. Like the Super-$Z$ factory, the GigaZ is a similar project \cite{LCCPhysicsWorkingGroup:2019fvj} based on the International Linear Collider (ILC) with a luminosity ${\cal L} \simeq 0.7\times10^{34}~{\rm cm}^{-2} {\rm s}^{-1}$. The Super-$Z$ factory can provide a great opportunity to detect and study some important particle states, like the doubly heavy baryons ($\Xi_{cc}$, $\Xi_{bc}$, $\Xi_{bb}$) and its excited states. Furthermore, it can also be utilized for the study of heavy flavor physics and the hadronization of QCD, which is of great significance to the development of particle physics. Several recent studies \cite{Zheng:2015ixa,Jiang:2012jt,Jiang:2013ej,Liao:2018nab} have illustrated the potential for the production and decay of the doubly heavy baryons and its excited states at the future Super-$Z$ factory. 

For the production of doubly heavy baryons at $e^{+} e^{-}$ collider, the typical physical process is $e^{+} + e^{-} \rightarrow \gamma^* / Z^* \rightarrow \Xi_{QQ^{\prime}} +\bar{Q^{\prime}} +\bar{Q}$, where $\Xi_{QQ^{\prime}}$ is used to represent the doubly heavy baryon without the light quark specified because the isospin-breaking effect of the doubly heavy baryons can be ignored in the manuscript. At the Super-$Z$ factory, the contribution from the intermediate $\gamma$ propagator is negligible. Thus we would focus on the production of doubly heavy baryons through the $Z^0$ propagator. Through $e^+ e^-$ annihilation, the production of doubly heavy baryon with intermediate $S$-wave ($[^1S_0]$ and $[^3S_1]$) diquark states have been analyzed \cite{Zheng:2015ixa,Jiang:2012jt,Jiang:2013ej}. Subsequently in this manuscript, we focus primarily on the production of excited doubly heavy baryon with $P$-wave diquark states, namely $\langle cc\rangle[^{1}P_{1}]_{\mathbf{\bar 3}}$, $\langle cc\rangle[^{3}P_{J}]_{\mathbf{6}}$, 
$\langle bc\rangle[^{1}P_{1}]_{\mathbf{\bar 3}/ \mathbf{6}}$, $\langle bc\rangle[^{3}P_{J}]_{\mathbf{\bar 3}/ \mathbf{6}}$, 
$\langle bb \rangle[^{1}P_{1}]_{\mathbf{\bar 3}}$, and $\langle bb\rangle[^{3}P_{J}]_{\mathbf{6}}$ with $J=0$, 1, or 2. The contribution of $P$-wave can serve as a correction of the dominant $S$-wave contribution. The doubly heavy baryon produced from the intermediate diquark states with higher spin ($[^3S_1]$) or higher orbital ($[^1P_1]$, $[^3P_J]$) quantum number can be regarded as the excited states of the doubly heavy baryon. These excited states could decay into the ground states ($[^1S_0]$) 100\% via electromagnetic or hadronic interactions. In addition to the total cross sections,  the differential distributions including the transverse momentum, rapidity, angular, and invariant mass, are also provided to present the detailed kinematic characteristics of doubly heavy baryons, which would also be useful to the experiment. 

The remaining parts of this manuscript are arranged as follows. In Sec.~\ref{method}, we give the detailed calculation technology for the production of excited doubly heavy baryons at the Super-$Z$ factory within the NRQCD framework. Then the numerical analysis of the total cross sections, the differential distributions, as well as the theoretical uncertainty are given in Sec.~\ref{result}. Finally, Sec.~\ref{summary} is reserved for a summary.

\section{Calculation Technology}\label{method}

In this section, the detailed calculation technology for the production of doubly heavy baryon associated with two antiquarks at the Super-$Z$ factory would be given based on the NRQCD factorization theorem. The total cross section of this process can be factorized into the following form:

\begin{eqnarray}
\sigma(e^{+} e^{-}\rightarrow \Xi_{QQ^{\prime}}+\bar{Q^{\prime}}+\bar{Q}) =\sum\limits_{n} \hat\sigma \left(e^{+} e^{-}\rightarrow \langle QQ^{\prime}\rangle[n]+\bar{Q^{\prime}}+\bar{Q}\right) \langle{\cal O}^H[n]\rangle, \label{total}
\end{eqnarray}
where $\langle{\cal O}^H[n]\rangle$ is the non-perturbative matrix element, which is proportional to the transition probability of the diquark $\langle QQ^{\prime}\rangle[n]$ into the doubly heavy baryon $\Xi_{QQ^{\prime}}$. For the diquark in color-antitriplet $\bar{\mathbf{3}}$, the transition probability can be obtained from the original Schrödinger wave function for the $S$-wave and its first derivative for the $P$-wave, which can be naturally connected to the radial wave function at the origin and its first derivative,
\begin{eqnarray}
\langle{\cal O}^H[S]_\mathbf{\bar{3}}\rangle=|\Psi(0)|^2=\frac{1}{4\pi}|R(0)|^2 , \nonumber\\
\langle{\cal O}^{H}[P]_\mathbf{\bar{3}}\rangle=|\Psi'(0)|^2=\frac{3}{4\pi}|R'(0)|^2.
\end{eqnarray}
However the transition probability of diquark in color-sextuplet $\mathbf 6$ is currently uncertain, and here we use the same transition probability as color-antitriplet $\bar{\mathbf 3}$.

For the $\langle cc\rangle[n]$ and $\langle bb\rangle[n]$ diquark states production in $S$-wave, there are two spin and color configurations: $[^1S_0]_{\bf 6}$ and $[^3S_1]_{\bf\bar{3}}$, while for in $P$-wave, there are four configurations: $[^{1}P_{1}]_{\mathbf{\bar 3}}$, $[^{3}P_{J}]_{\mathbf{6}}$ ($J=0, 1, 2$). As for the $(bc)$-diquark, there are four spin and color configurations for $S$-wave: $[^1S_0]_{\bf\bar{3}/\bf 6}$ and $[^3S_1]_{\bf\bar{3}/\bf 6}$, and eight configurations for $P$-wave: $\langle bc\rangle[^{1}P_{1}]_{\mathbf{\bar 3}/ \mathbf{6}}$, $\langle bc\rangle[^{3}P_{J}]_{\mathbf{\bar 3}/ \mathbf{6}}$ ($J=0,~1$, or 2).

\begin{figure}[htb]
  \centering
  \subfigure[]{
    \includegraphics[scale=0.33]{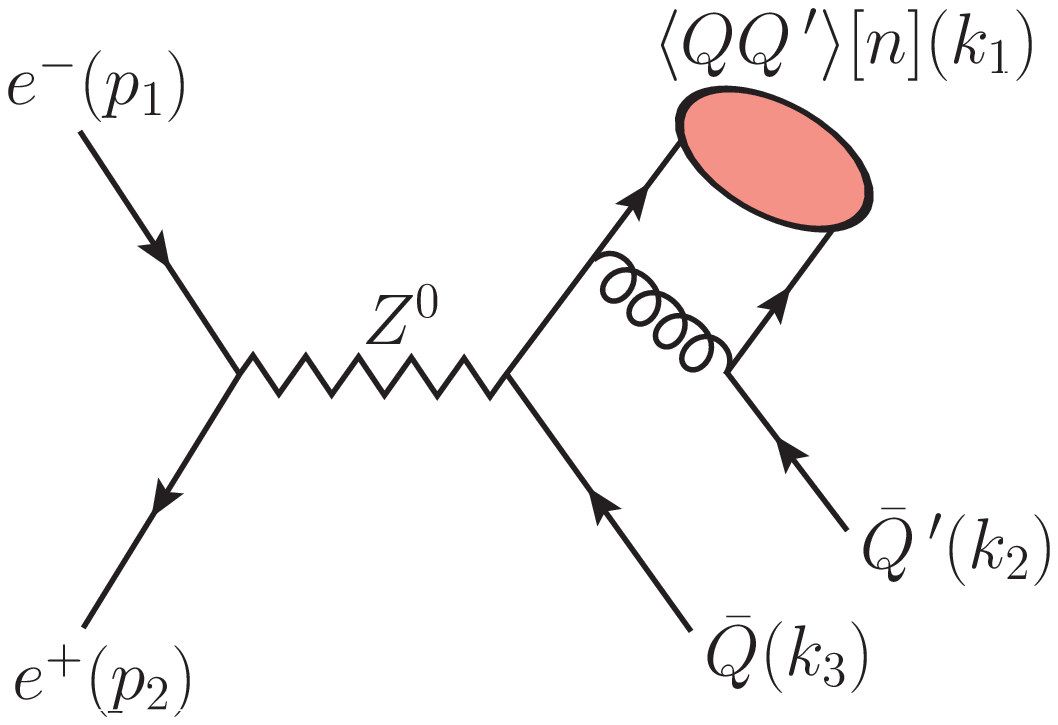}}
  \hspace{0.00in}
  \subfigure[]{
    \includegraphics[scale=0.33]{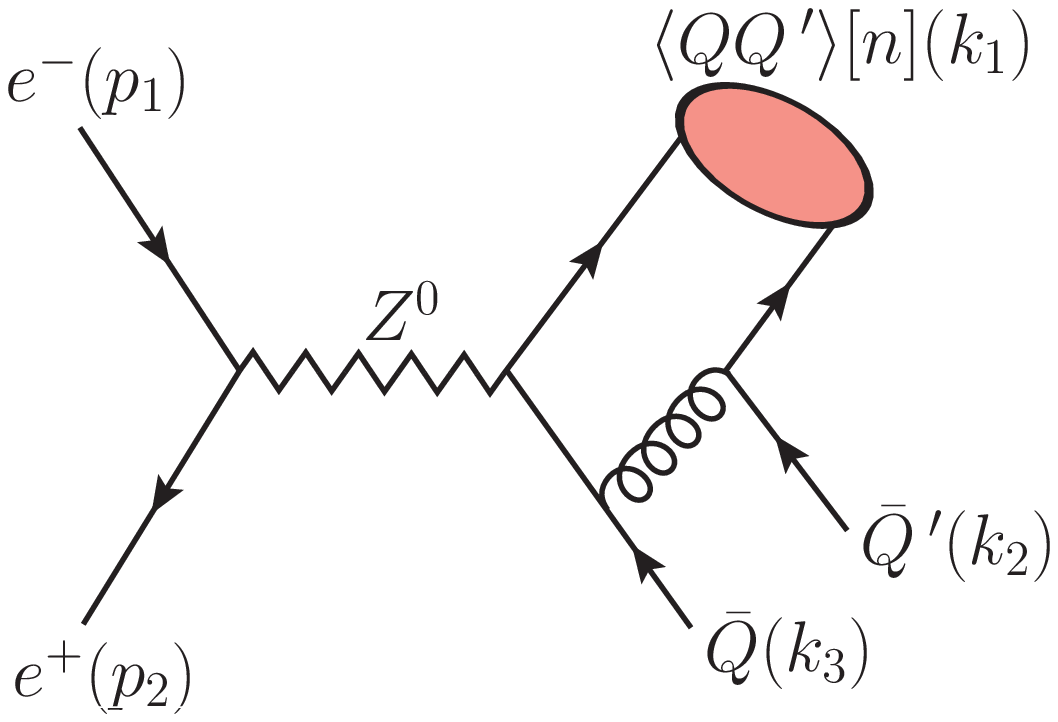}}
  \hspace{0.00in}
  \subfigure[]{
    \includegraphics[scale=0.33]{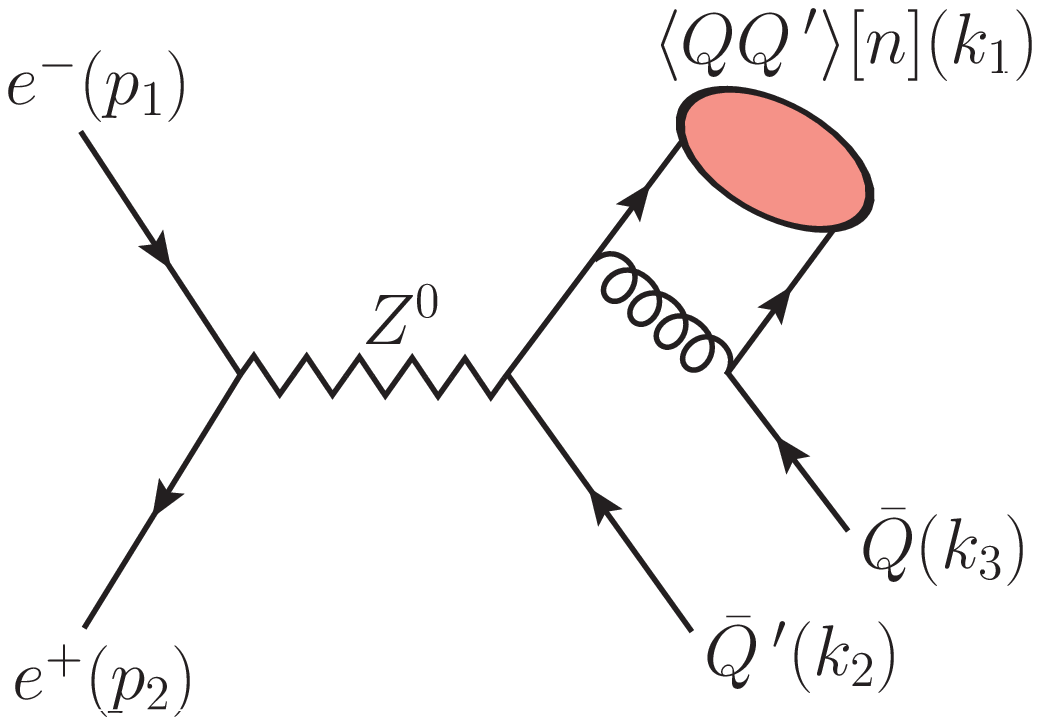}}
  \hspace{0.00in}
  \subfigure[]{
    \includegraphics[scale=0.33]{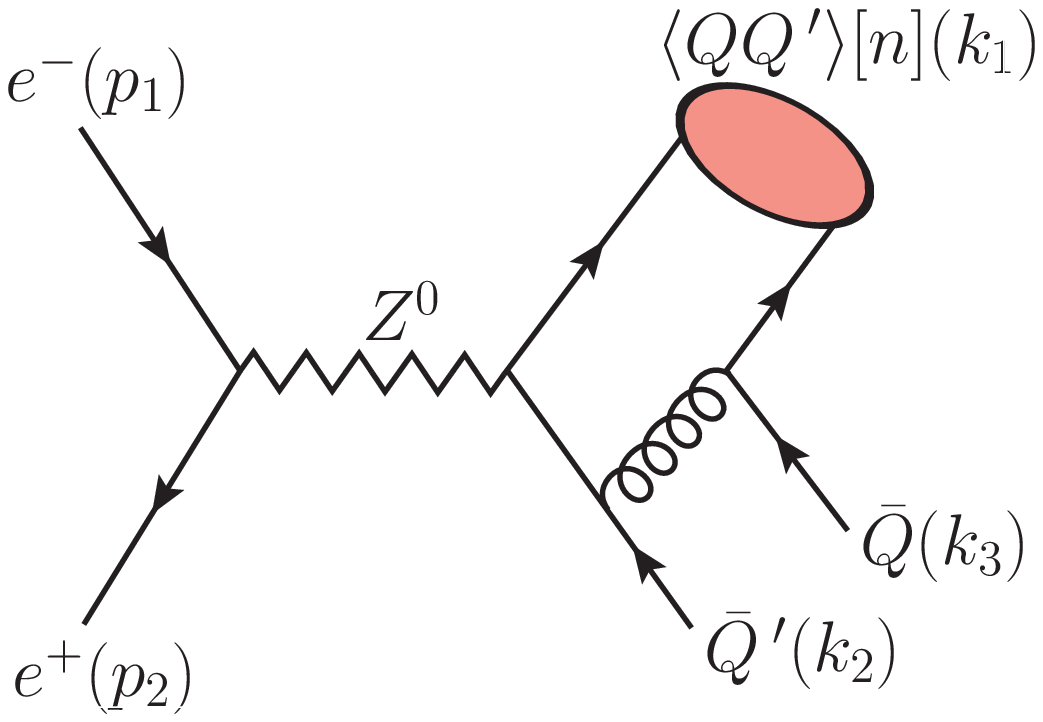}}
  \caption{Feynman diagrams for the process $e^{-}(p_1) + e^{+}(p_2) \rightarrow \langle QQ^{\prime}\rangle[n](k_1)+ \bar {Q^{\prime}} (k_2) + \bar{Q} (k_3)$, where $Q^{(\prime)}$ denotes as heavy $c$ or $b$ quark in $\langle QQ'\rangle[n]$, i.e., $\langle cc\rangle[^{1}P_{1}]_{\mathbf{\bar 3}}$, $\langle cc\rangle[^{3}P_{J}]_{\mathbf{6}}$, 
$\langle bc\rangle[^{1}P_{1}]_{\mathbf{\bar 3}/ \mathbf{6}}$, $\langle bc\rangle[^{3}P_{J}]_{\mathbf{\bar 3}/ \mathbf{6}}$, 
$\langle bb \rangle[^{1}P_{1}]_{\mathbf{\bar 3}}$ and $\langle bb\rangle[^{3}P_{J}]_{\mathbf{6}}$, with $J=0,~1$, or 2 for the $P$-wave diquark.}
  \label{diagram1}
\end{figure}

At the Super-$Z$ factory, we focus on the production of doubly heavy baryon through the $Z^0$  exchange, as the contribution of the intermediate $\gamma$ propagator is relatively small. The corresponding Feynman diagrams for this perturbative process $e^{-}(p_1) + e^{+}(p_2) \rightarrow Z^0 \rightarrow \langle QQ^{\prime}\rangle[n](k_1) +\bar{Q^{\prime}}(k_2) +\bar{Q}(k_3)$ are shown in Fig.~\ref{diagram1}. To be mentioned, for the production of $\langle cc\rangle$ and $\langle bb\rangle$ diquarks, there are four additional Feynman diagrams by exchanging the two identical heavy quark line in the diquark.
The perturbative differential cross section in Eq.~\ref{total} can be written as 
\begin{eqnarray}
d\hat\sigma\left(e^{+} e^{-}\rightarrow \langle QQ^{\prime}\rangle[n]+\bar{Q^{\prime}}+\bar{Q}\right)  = \frac{\overline{\sum} |{\mathcal{M}[n]}|^{2} d\Phi_3}{4\sqrt{(p_1\cdot p_2)^2-m_e^4}} , \label{cs}
\end{eqnarray}
where $m_e$ denotes the mass of electron and positron, $\overline{\sum}$ means to average over the spin states of the electron and positron, and sum over the color and spin of the final-state particles. $d{\Phi_3}$ represents the 3-body phase space, $\mathcal{M}[n]$ is the hard amplitude for the production of diquark $\langle QQ^{\prime}\rangle$ with spin and color quantum number $[n]$. For the production of diquark $\langle QQ^{\prime}\rangle[n]$ in $S$-wave, the fermion line $Q^{\prime}\bar{Q^{\prime}}$ needs to be reversed by properly inserting the charge conjugate matrix $C = -i\gamma^2 \gamma^0$ into the amplitudes. After this action, the amplitudes can be related to the familiar meson production $( Q\bar {Q^{\prime}})[n](k_1)+ Q^{\prime} (k_2) + \bar{Q} (k_3)$ with an additional factor $(-1)^{\zeta +1}$, where $\zeta$ stands for the number of vector vertices in the $Q^{\prime}\bar{Q^{\prime}}$ fermion line. Specifically, $\zeta = 1$ for the subfigures $(a)$ and $(b)$ in Fig.~\ref{diagram1}; as for the last two subfigures $(c)$ and $(d)$, $\zeta = 2$ ($\zeta = 1$) for vector (axial-vector) contributions of the interaction vertex $\Gamma_{ZQ^{\prime}}$. A detailed proof can be found in Ref.~\cite{Zheng:2015ixa,Jiang:2012jt}.

We now present the hard amplitude $\mathcal M[n]$ of the $P$-wave diquark states, which can be obtained by taking the first derivative of the relative momentum $k$ between these two constituent quarks of the diquark in the amplitude of $S$-wave. The relative momentum $k$ appears primarily as the constituent quark momenta of the diquark, $k_{11}=\frac{m_Q}{M_{QQ^{\prime}}}k_1+k$ and $k_{12}=\frac{m_{Q^{\prime}}}{M_{QQ^{\prime}}}k_1-k$, in the projector and propagators. Here, ${m_Q}$ and ${m_{Q^{\prime}}}$ are the masses of heavy quarks. To ensure the gauge invariance, the mass of diquark $M_{QQ'}$ is taken to be $m_{Q}+ m_{Q'}$. Specifically, the amplitude $\mathcal{M}[P]$ can be written as 

\begin{widetext}
\begin{eqnarray}\label{m1p11}
\mathcal{M}_{a}[^1P_1]&=& \kappa \left. \varepsilon^l_{\alpha}(k_1)\frac{d}{dk_{\alpha}}\left[\mathcal{L}^\nu_{ss^{\prime}}\mathcal{D}_{\mu\nu} \bar{u}_i(k_2) \gamma_{\rho}  \frac{\Pi_{[^1 S_0]}(k_1)}{(q_2+k_{12})^2} \gamma_{\rho} \frac{\slashed{k}_{1}+\slashed{k}_{2}+m_{Q}}{(k_1+k_2)^2-m_{Q}^{2}}\Gamma^\mu_{ZQ}  v_j(k_3)\right]\right|_{k=0}, \\
\mathcal{M}_{b}[^1P_1]&=& \kappa  \left. \varepsilon^l_{\alpha}(k_1)\frac{d}{dk_{\alpha}}\left[\mathcal{L}^\nu_{ss^{\prime}}\mathcal{D}_{\mu\nu} \bar{u}_i(k_2) \gamma_{\rho}  \frac{\Pi_{[^1 S_0]}(k_1)}{(k_2+k_{12})^2} \Gamma^\mu_{ZQ} \frac{-\slashed{k}_{12}-\slashed{k}_{2}-\slashed{k}_{3}+m_{Q}}{(k_{12}+k_2+k_3)^2-m_{Q}^{2}} \gamma_{\rho} v_j(k_3)\right]\right|_{k=0},  \\
\mathcal{M}_{c}[^1P_1]&=& \kappa \left. \varepsilon^l_{\alpha}(k_1)\frac{d}{dk_{\alpha}}\left[\mathcal{L}^\nu_{ss^{\prime}}\mathcal{D}_{\mu\nu}\bar{u}_i(k_2) \gamma_\rho \frac{\slashed{k}_{11}+\slashed{k}_{2}+\slashed{k}_{3}+m_{Q^{\prime}}}{(k_{11}+k_2+k_3)^2-m_{Q^{\prime}}^{2}}
\Gamma_{ZQ^{\prime}}^{\mu} \frac{\Pi_{[^1 S_0]}(k_1)}{(k_3+k_{11})^2} \gamma_{\rho} v_j(k_3)\right]\right|_{k=0}, \\
\mathcal{M}_{d}[^1P_1]&=& \kappa \left. \varepsilon^l_{\alpha}(k_1)\frac{d}{dk_{\alpha}}\left[\mathcal{L}^\nu_{ss^{\prime}}\mathcal{D}_{\mu\nu} \bar{u}_i(k_2) \Gamma^\mu_{ZQ^{\prime}} \frac{-\slashed{k}_{1}-\slashed{k}_{3}+m_{Q^{\prime}}}{(k_{1}+k_3)^2-m_{Q^{\prime}}^{2}} \gamma_{\rho} \frac{\Pi_{[^1 S_0]}(k_1)}{(k_3+k_{11})^2}  \gamma_{\rho} v_j(k_3)\right]\right|_{k=0},\label{m1p14}
\end{eqnarray}
\end{widetext}
and
\begin{widetext}
\begin{eqnarray}\label{m3pj1}
\mathcal{M}_{a}[^3P_J]&=& \kappa \left. \varepsilon^J_{\alpha \beta}(k_1)\frac{d}{dk_{\alpha}}\left[ \mathcal{L}^\nu_{ss^{\prime}}\mathcal{D}_{\mu\nu} \bar{u}_i(k_2) \gamma_{\rho}  \frac{\Pi_{[^3 S_1]}^{\beta}(k_1)}{(q_2+k_{12})^2} \gamma_{\rho} \frac{\slashed{k}_{1}+\slashed{k}_{2}+m_{Q}}{(k_1+k_2)^2-m_{Q}^{2}}\Gamma^\mu_{ZQ}  v_j(k_3)\right]\right|_{k=0}, \\
\mathcal{M}_{b}[^3P_J]&=& \kappa \left. \varepsilon^J_{\alpha \beta}(k_1)\frac{d}{dk_{\alpha}}\left[ \mathcal{L}^\nu_{ss^{\prime}}\mathcal{D}_{\mu\nu} \bar{u}_i(k_2) \gamma_{\rho}  \frac{\Pi_{[^3 S_1]}^{\beta}(k_1)}{(k_2+k_{12})^2} \Gamma^\mu_{ZQ} \frac{-\slashed{k}_{12}-\slashed{k}_{2}-\slashed{k}_{3}+m_{Q}}{(k_{12}+k_2+k_3)^2-m_{Q}^{2}} \gamma_{\rho} v_j(k_3)\right]\right|_{k=0},  \\
\mathcal{M}_{c}[^3P_J]&=& \kappa \left. \varepsilon^J_{\alpha \beta}(k_1)\frac{d}{dk_{\alpha}}\left[\mathcal{L}^\nu_{ss^{\prime}}\mathcal{D}_{\mu\nu}\bar{u}_i(k_2) \gamma_\rho \frac{\slashed{k}_{11}+\slashed{k}_{2}+\slashed{k}_{3}+m_{Q^{\prime}}}{(k_{11}+k_2+k_3)^2-m_{Q^{\prime}}^{2}}
\Gamma_{ZQ^{\prime}}^{\mu} \frac{\Pi_{[^3 S_1]}^{\beta}(k_1)}{(k_3+k_{11})^2} \gamma_{\rho} v_j(k_3)\right]\right|_{k=0},\\
\mathcal{M}_{d}[^3P_J]&=& \kappa \left. \varepsilon^J_{\alpha \beta}(k_1)\frac{d}{dk_{\alpha}}\left[\mathcal{L}^\nu_{ss^{\prime}}\mathcal{D}_{\mu\nu} \bar{u}_i(k_2) \Gamma^\mu_{ZQ^{\prime}} \frac{-\slashed{k}_{1}-\slashed{k}_{3}+m_{Q^{\prime}}}{(k_{1}+k_3)^2-m_{Q^{\prime}}^{2}} \gamma_{\rho} \frac{\Pi_{[^3 S_1]}^{\beta}(k_1)}{(k_3+k_{11})^2}  \gamma_{\rho} v_j(k_3)\right]\right|_{k=0}, \label{m3pj4}
\end{eqnarray}
\end{widetext}
where the overall parameter $\kappa=\frac{\mathcal{C}g^2 g_s^2}{cos^2{\theta_W}}$, $\Gamma^{\mu}_{ZQ^{(\prime)}}$ means the interaction vertex,
\begin{eqnarray}\label{vertex}
 \Gamma^{\mu}_{Zc}&=&\gamma^{\mu} \left[\alpha \left( \frac{1}{4}-\frac{2}{3} \sin ^2 \theta_w \right)-\frac{\gamma^5}{4} \right], \\
 \Gamma^{\mu}_{Zb}&=&-\gamma^{\mu} \left[\alpha \left( \frac{1}{4}-\frac{1}{3} \sin ^2 \theta_w\right) - \frac{\gamma^5}{4} \right],
\end{eqnarray}
where $\alpha = 1$ for $Z^0-Q-\bar{Q}$ vertex and $\alpha = -1$ for the vertex $Z^0 - Q^{\prime}-\bar{Q}^{\prime}$,
the leptonic vector $\mathcal{L}^\nu_{ss^{\prime}}=\bar{\nu}_s\left(p_2\right) \gamma^\nu\left(\sin ^2 \theta_w+\frac{\gamma^5}{4} -\frac{1}{4}\right)$ $u_{s^{\prime}}\left(p_1\right)$, $Z^0$ propagator $\mathcal{D}_{\mu\nu}=\frac{i}{p^2-m_Z^2+im_Z\Gamma_Z}\left(-g_{\mu \nu}+\frac{p_\mu p_\nu}{p^2}\right)$, $\Gamma_Z$ is the total decay width of $Z^0$ boson, and $\theta_W$ refers to the Weinberg angle. $\varepsilon^l_{\alpha}$ is the polarization vector of $^1P_1$ diquark and $\varepsilon^J_{\alpha\beta}$ is the polarization tensor of $^3P_J$ diquark with $J=0,~1$, or 2. Both of them need to be polarization summed to select the proper total angular momentum, and the sum formulas of polarization vector and polarization tensor satisfy the following forms~\cite{Petrelli:1997ge},
\begin{eqnarray}
	\sum_{l_{z}} \varepsilon^l_{\alpha} \varepsilon_{\alpha^{\prime}}^{l*} &=&\Pi_{\alpha \alpha^{\prime}},\\
	\varepsilon_{\alpha \beta}^{0} \varepsilon_{\alpha^{\prime} \beta^{\prime}}^{0*} &=&\frac{1}{3} \Pi_{\alpha \beta} \Pi_{\alpha^{\prime} \beta^{\prime}}, \label{psum3p0}\\
	\sum_{J_{z}} \varepsilon_{\alpha \beta}^{1} \varepsilon_{\alpha^{\prime} \beta^{\prime}}^{1 *} &=&\frac{1}{2}\left(\Pi_{\alpha \alpha^{\prime}} \Pi_{\beta \beta^{\prime}}-\Pi_{\alpha \beta^{\prime}} \Pi_{\alpha^{\prime} \beta}\right), \label{psum3p1}\\
	\sum_{J_{z}} \varepsilon_{\alpha \beta}^{2} \varepsilon_{\alpha^{\prime} \beta^{\prime}}^{2 *} &=&\frac{1}{2}\left(\Pi_{\alpha \alpha^{\prime}} \Pi_{\beta \beta^{\prime}}+\Pi_{\alpha \beta^{\prime}} \Pi_{\alpha^{\prime} \beta}\right)-\frac{1}{3} \Pi_{\alpha \beta} \Pi_{\alpha^{\prime} \beta^{\prime}},\label{psum3p2}
\end{eqnarray}
with the definition 
\begin{eqnarray}
	\Pi_{\alpha \beta}=-g_{\alpha \beta}+\frac{p_{1 \alpha} p_{1 \beta}}{M_{QQ'}^{2}}.
\end{eqnarray}

The color factor $\mathcal{C}$ in overall parameter $\kappa$ can be described as
\begin{eqnarray}
\mathcal{C}_{ij,l}=\frac{1}{\sqrt{2}} \times \sum_{a=1}^{8}  \sum_{m,n=1}^{3} (T^a)_{mi} (T^a)_{nj} \times G_{mnl},
\end{eqnarray}
where $i, j, m, n$ are the color indices of the outgoing $\bar{Q}$, $\bar{Q'}$, $Q$ and $Q^{\prime}$, respectively, $a$ and $l$ denote the color indices of the gluon and the diquark $\langle QQ^{\prime}\rangle$; $G_{mnl}$ can be considered equal to the antisymmetric function $\varepsilon_{mnl}$ for $\bf \bar{3}$ state, or the symmetric function $f_{mnl}$ for $\bf 6$ state, which satisfy
\begin{eqnarray}
\varepsilon_{mnl} \varepsilon_{m^{\prime}n^{\prime}l}=\delta_{mm^{\prime}}\delta_{nn^{\prime}}-\delta_{mn^{\prime}}\delta_{nm^{\prime}},
\nonumber\\f_{mnl} f_{m^{\prime}n^{\prime}l}=\delta_{mm^{\prime}}\delta_{nn^{\prime}}+\delta_{mn^{\prime}}\delta_{nm^{\prime}}.
\end{eqnarray}

The projector $\Pi_{[^1 S_0]}(k_1)$, $\Pi_{[^3 S_1]}^{\beta}(k_1)$, and its first derivative of the relative momentum $k$ are
\begin{widetext}
\begin{eqnarray}
\Pi_{[^1 S_0]}(k_1) &=& \frac{-\sqrt{M_{QQ^{\prime}}}}{4 m_{Q} m_{Q^{\prime}}}\left(\slashed{k}_{12}-m_{Q^{\prime}}\right) \gamma^{5}\left(\slashed{k}_{11}+m_{Q}\right),\\
\Pi_{[^3 S_1]}^{\beta}(k_1)&=& \frac{-\sqrt{M_{QQ^{\prime}}}}{4 m_{Q} m_{Q^{\prime}}}\left(\slashed{k}_{12}-m_{Q^{\prime}}\right) \gamma^{\beta} \left(\slashed{k}_{11}+m_{Q}\right),\\
\left. \frac{d}{dk_{\alpha}}\Pi_{[^1 S_0]}(k_1) \right|_{k=0}&=& \frac{\sqrt{M_{QQ^{\prime}}}}{4 m_{Q} m_{Q^{\prime}}}\gamma^{\alpha} \gamma^{5} \left(\slashed{k}_{1}+m_{Q}-m_{Q^{\prime}}\right), \\
\left. \frac{d}{dk_{\alpha}}\Pi_{[^3 S_1]}^{\beta}(k_1) \right|_{k=0}&=& \frac{\sqrt{M_{QQ^{\prime}}}}{4 m_{Q} m_{Q^{\prime}}} \left[ \gamma^{\alpha} \gamma^{\beta} \left(\slashed{k}_{1}+m_{Q}-m_{Q^{\prime}}\right) -2 g^{\alpha \beta} \left(\slashed{k}_{12} - m_{Q^{\prime}} \right) \right].
\end{eqnarray}
\end{widetext}

\section{Numerical analysis}\label{result}

The input parameters used in the numerical calculation are taken as follows~\cite{Baranov:1995rc,Patrignani:2016xqp,Kiselev:2002iy}: 
\begin{eqnarray}
G_F=1.16637 \times 10^{-5}~\rm{GeV}^{-2}, \Gamma_z&=&2.4952~\rm{GeV},~\it m_Z=\rm 91.1876~GeV, \nonumber\\
m_W=80.385~\rm{GeV},~\it m_c&=&\rm 1.8~ GeV,~\it m_b=\rm 5.1~GeV,\nonumber\\
M_{\Xi_{cc}}=3.6~{\rm GeV}, M_{\Xi_{bc}}&=&6.9~{\rm GeV}, M_{\Xi_{bb}}=10.2~{\rm GeV},\nonumber\\
|R_{cc}(0)|=~{\rm 0.700~GeV}^{3/2}, |R_{bc}(0)|&=&0.904~{\rm GeV}^{3/2}, |R_{bb}(0)|=1.382~{\rm GeV}^{3/2},\nonumber\\
|R'_{cc}(0)|=~{\rm 0.102~GeV}^{5/2}, |R'_{bc}(0)|&=&0.200~{\rm GeV}^{5/2}, |R'_{bb}(0)|=0.479~{\rm GeV}^{5/2}.\nonumber
\end{eqnarray}
The strong coupling constant is set to be $\alpha_s( 2m_c)=\rm 0.242$ for $\Xi_{cc}$ and $\Xi_{bc}$ production, and $\alpha_s( 2 m_b)=\rm 0.180$ for the production of $\Xi_{bb}$.

Because of the identity for the  $\langle cc\rangle$ and $\langle bb\rangle$ diquark production, an additional factor of $2^2/(2!2!)$ should be multiplied, where $2^2$ contributes to the four additional Feynman diagrams, and $1/(2!2!)$ accounts for the identities of two heavy quarks inside the diquark and two final-state antiquarks.
The color factor $\mathcal{C}^{2}=4/3$ for the production of color $\bf \bar{3}$ state, and $2/3$ for $\bf 6$ state.

With the help of FeynArts 3.9~\cite{Hahn:2000kx}, FeyCalc 9.3~\cite{Shtabovenko:2020gxv}and modified FormCalc~\cite{Hahn:1998yk} programs, the cross sections and the differential distributions of all considered intermediate diquark states can be obtained.
The cross sections ($\sigma$) and events ($N$) for the excited $\Xi_{cc}$, $\Xi_{bc}$, and $\Xi_{bb}$ baryons produced at the Super-$Z$ factory ($\sqrt{S}=m_Z$) with luminosity ${\cal L} \simeq 10^{34}~{\rm cm}^{-2} {\rm s}^{-1}$ are presented in Tables \ref{csccbb} and \ref{csbc}. The events produced at the GigaZ ($N_{\rm GigaZ}$) with luminosity ${\cal L} \simeq 0.7\times10^{34}~{\rm cm}^{-2} {\rm s}^{-1}$ are also displayed in Tables \ref{csccbb} and \ref{csbc}. In order to compare with the contribution from $S$-wave diquark states, the corresponding cross sections and events of $S$-wave diquarks produced at the Super-$Z$ factory and GigaZ are also presented in Tables \ref{csccbb} and \ref{csbc}. Our estimation for the $\Xi_{QQ^{\prime}}$ production with $S$-wave diquark states at Super-$Z$ factories is consistent with literature \cite{Zheng:2015ixa,Ali:2018ifm,Qin:2020zlg}. However, they are slightly different from those in the previous literature \cite{Jiang:2013ej} for the reason that we revised the contribution of the axial vector in the interaction vertex $\Gamma_{ZQ^{\prime}}^{\mu}$ on the reversed fermion line  $Q^{\prime}\bar Q^{\prime}$ with an extra factor -1 in subfigures $(c)$ and $(d)$ of Fig. \ref{diagram1}.

\black
\begin{table}[htb]
\caption{Cross sections (in unit: fb) and events (in unit: $10^{2}$) for $\Xi_{cc}$ and $\Xi_{bb}$ baryons produced at the Super-$Z$ factory and GigaZ, respectively. }
\centering
\renewcommand\arraystretch{1.2}
\resizebox{\linewidth}{!}{
\begin{tabular}{|c||c|c|c|c|c|c||c|c|c|c|c|c|}
\hline
\multirow{2}*{State} & \multicolumn{6}{|c||}{$\Xi_{cc}$} & \multicolumn{6}{c|}{$\Xi_{bb}$} \\
\cline{2-13}
~& $[^1S_0]_{\mathbf{6}}$ & $[^3S_1]_{\overline{\mathbf{3}}}$ & $[^1P_1]_{\overline{\mathbf{3}}}$ & $[^3P_0]_{\mathbf{6}}$ & $[^3P_1]_{\mathbf{6}}$ & $[^3P_2]_{\mathbf{6}}$ & $[^1S_0]_{\mathbf{6}}$ & $[^3S_1]_{\overline{\mathbf{3}}}$ & $[^1P_1]_{\overline{\mathbf{3}}}$ & $[^3P_0]_{\mathbf{6}}$ & $[^3P_1]_{\mathbf{6}}$ & $[^3P_2]_{\mathbf{6}}$ \\
\hline\hline
 $\sigma$   & 267.02 & 548.63 & 11.43  & 8.23  & 9.14  & 3.58 & 25.94 & 12.93 & 0.73  & 0.61  & 0.69  & 0.27  \\
\hline
 $N$  & 267.02 & 548.63 & 11.43  & 8.23  & 9.14  & 3.58  & 25.94 & 12.93 & 0.73  & 0.61  & 0.69  & 0.27 \\
\hline
$N_{\rm GigaZ}$  & 186.91 & 384.04 & 8.00  & 5.76  & 6.40  & 2.51  & 18.16 & 9.05 & 0.51  & 0.43  & 0.48  & 0.19 \\
\hline
\end{tabular}
}\label{csccbb}
\end{table}

\begin{table}[htb]
\caption{Cross sections (in unit: fb) and events (in unit: $10^{2}$) for $\Xi_{bc}$ baryon produced at the Super-$Z$ factory and GigaZ, respectively. }
\centering
\renewcommand\arraystretch{1.2}
\resizebox{\linewidth}{!}{
\begin{tabular}{|c||c|c|c|c|c|c|c|c|c|c|c|c|}
\hline
\multirow{2}*{State} & \multicolumn{12}{|c|}{$\Xi_{bc}$} \\
\cline{2-13}
~&  $[^1S_0]_{\overline{\mathbf{3}}}$ & $[^1S_0]_{\mathbf{6}}$ & $[^3S_1]_{\overline{\mathbf{3}}}$ & $[^3S_1]_{\mathbf{6}}$ & $[^1P_1]_{\overline{\mathbf{3}}}$ & $[^1P_1]_{\mathbf{6}}$ & $[^3P_0]_{\overline{\mathbf{3}}}$ & $[^3P_0]_{\mathbf{6}}$ & $[^3P_1]_{\overline{\mathbf{3}}}$ & $[^3P_1]_{\mathbf{6}}$ & $[^3P_2]_{\overline{\mathbf{3}}}$ & $[^3P_2]_{\mathbf{6}}$ \\
\hline\hline
 $\sigma$  & 609.10 & 304.55 & 825.00 & 412.50 & 17.28  & 8.64  & 11.95  & 5.98  & 22.24  & 11.12  & 21.43  & 10.72  \\
\hline
 $N$ & 609.10 & 304.55 & 825.00 & 412.50 & 17.28  & 8.64 & 11.95  & 5.98 & 22.24  & 11.12  & 21.43  & 10.72  \\
\hline
$N_{\rm GigaZ}$  & 426.37 & 213.19 & 577.50 & 288.75 & 12.09  & 6.05 & 8.37 & 4.18 & 15.57  & 7.79 & 15.00  & 7.50 \\
\hline
\end{tabular}
}\label{csbc}
\end{table}

Tables \ref{csccbb} and \ref{csbc} reveal that:
\begin{itemize}
\item By summing over all the different $P$-wave configurations of $\langle QQ^{\prime}\rangle$ diquark, the total cross sections of excited $\Xi_{cc}$, $\Xi_{bc}$, and $\Xi_{bb}$ from the orbital excited diquark states are $32.38$ fb, $109.36$ fb, and $2.29$ fb, respectively. These contributions correspond to 3.97 $\%$, 5.08 $\%$, and 5.89 $\%$ of the contributions from the total $S$-wave.
\item The largest $P$-wave contribution arises from the $[^3P_1]_{\overline{\mathbf{3}}}$ intermediate diquark state for $\Xi_{bc}$ production, and $[^1P_1]_{\overline{\mathbf{3}}}$ for $\Xi_{cc}$ and $\Xi_{bb}$ production.
\item At the Super-$Z$ factory, the total produced events of the excited $\Xi_{cc}$, $\Xi_{bc}$, and $\Xi_{bb}$ from all summed $P$-wave diquark states are $3.24 \times10^{3}$, $1.09 \times10^{4}$, and $2.30 \times10^{2}$, respectively, per year. While at the GigaZ, the total produced events per year are $2.27 \times10^{3}$, $7.66 \times10^{3}$, and $1.60 \times10^{2}$, accordingly.
\item Assuming that all the considered excited states can decay into the ground state 100 $\%$, the total cross sections for $\Xi_{cc}$, $\Xi_{bc}$, and $\Xi_{bb}$ production are $848.03$ fb, $2260.51$ fb, and $41.16$ fb, resulting in a large number of produced events up to $8.48 \times10^{4}$, $2.26 \times10^{5}$, and $4.12 \times10^{3}$. 
\item When the luminosity of the Super-$Z$ factory increases to $10^{36}~{\rm cm}^{-2} {\rm s}^{-1}$, the corresponding produced events of doubly heavy baryons will increase by 100 times. Considering the subsequent decay of $\Xi_{QQ^{\prime}}$ and the abilities of the detector, they are likely to be discovered experimentally at the future Super-$Z$ factory.
\end{itemize}

\begin{figure}[htb]
  \centering
  \hspace{-0.7in}
  \includegraphics[width=2.8in]{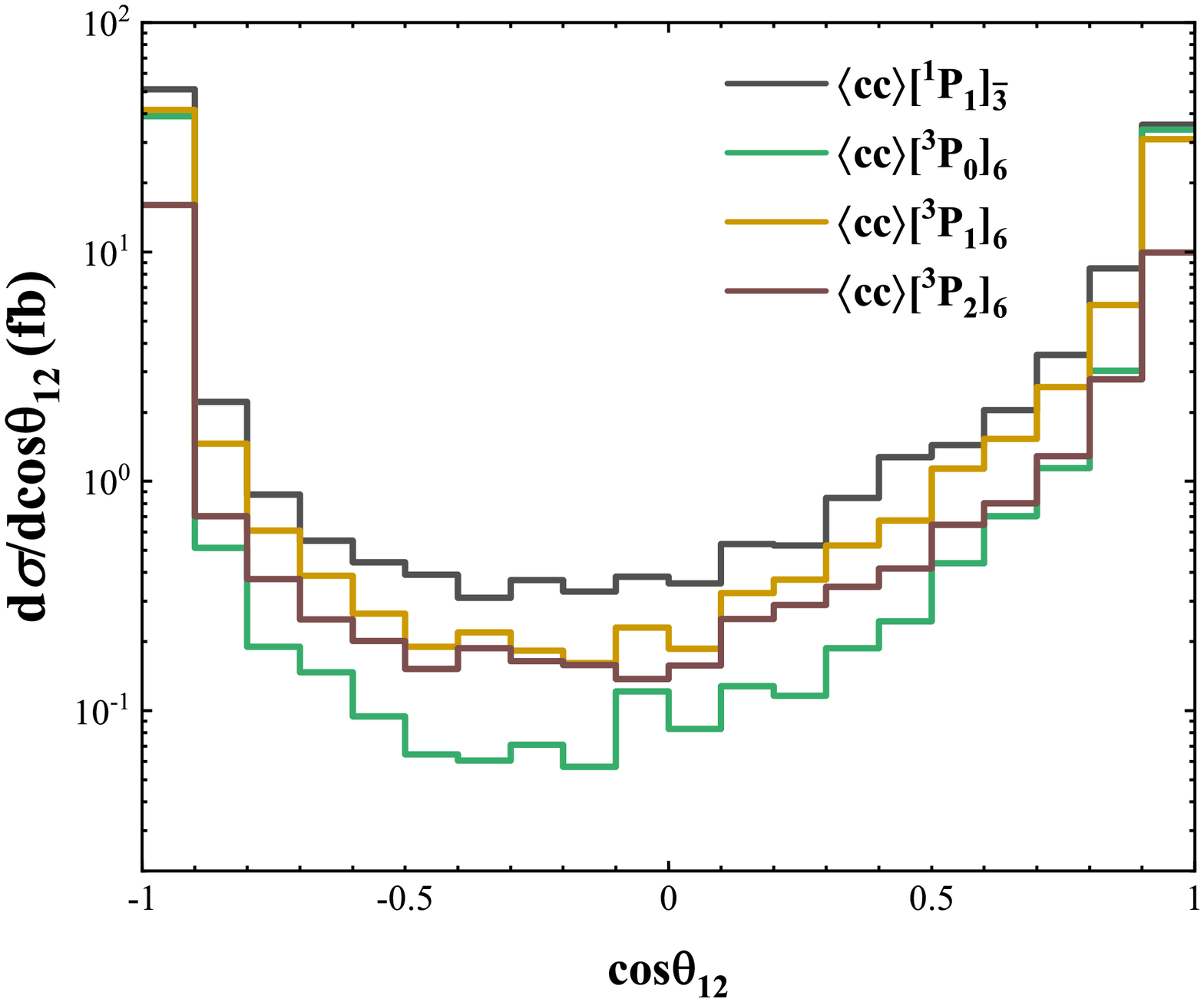}
  \hspace{-0.7in}
    \includegraphics[width=2.8in]{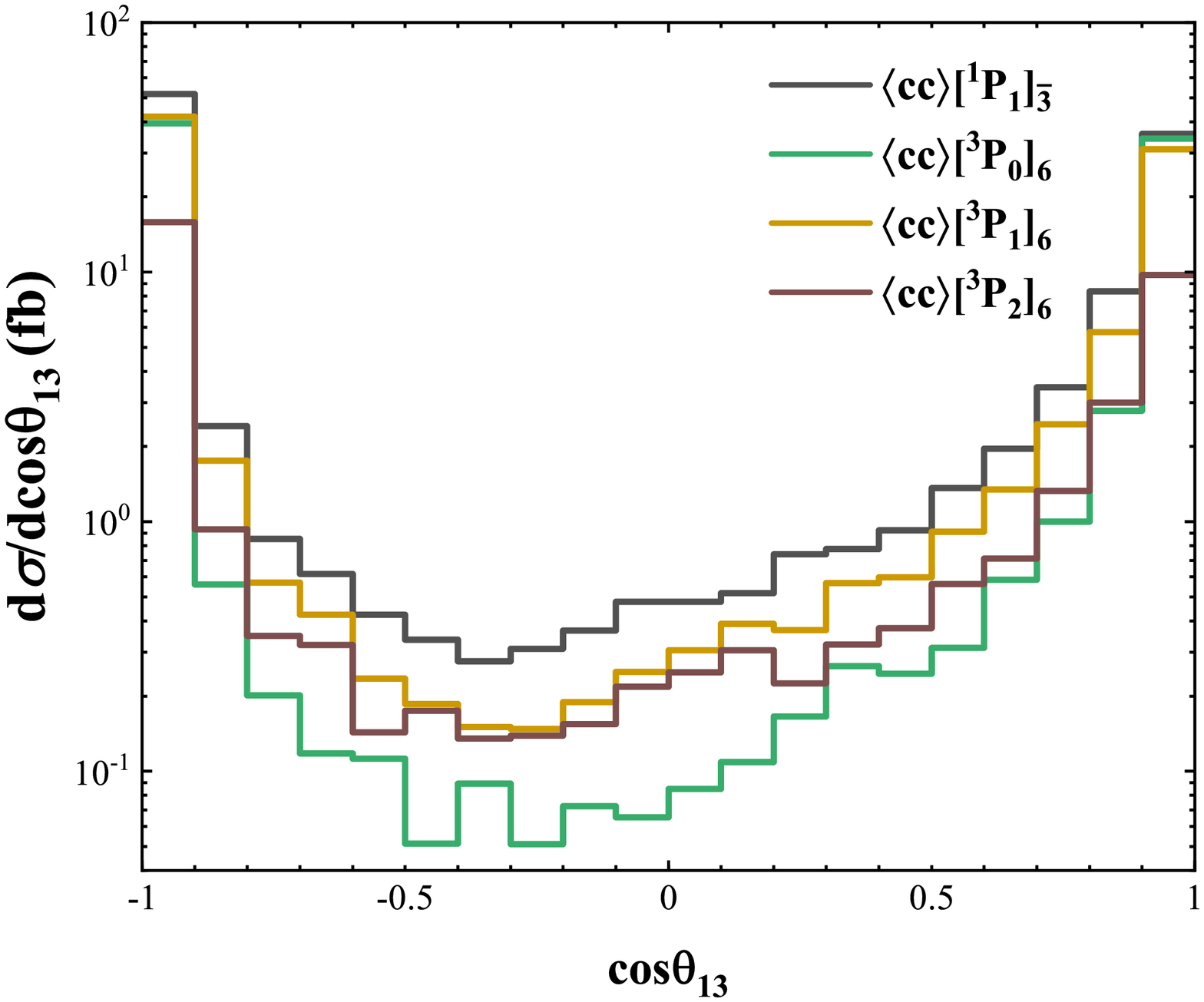}
  \hspace{-0.7in}
    \includegraphics[width=2.8in]{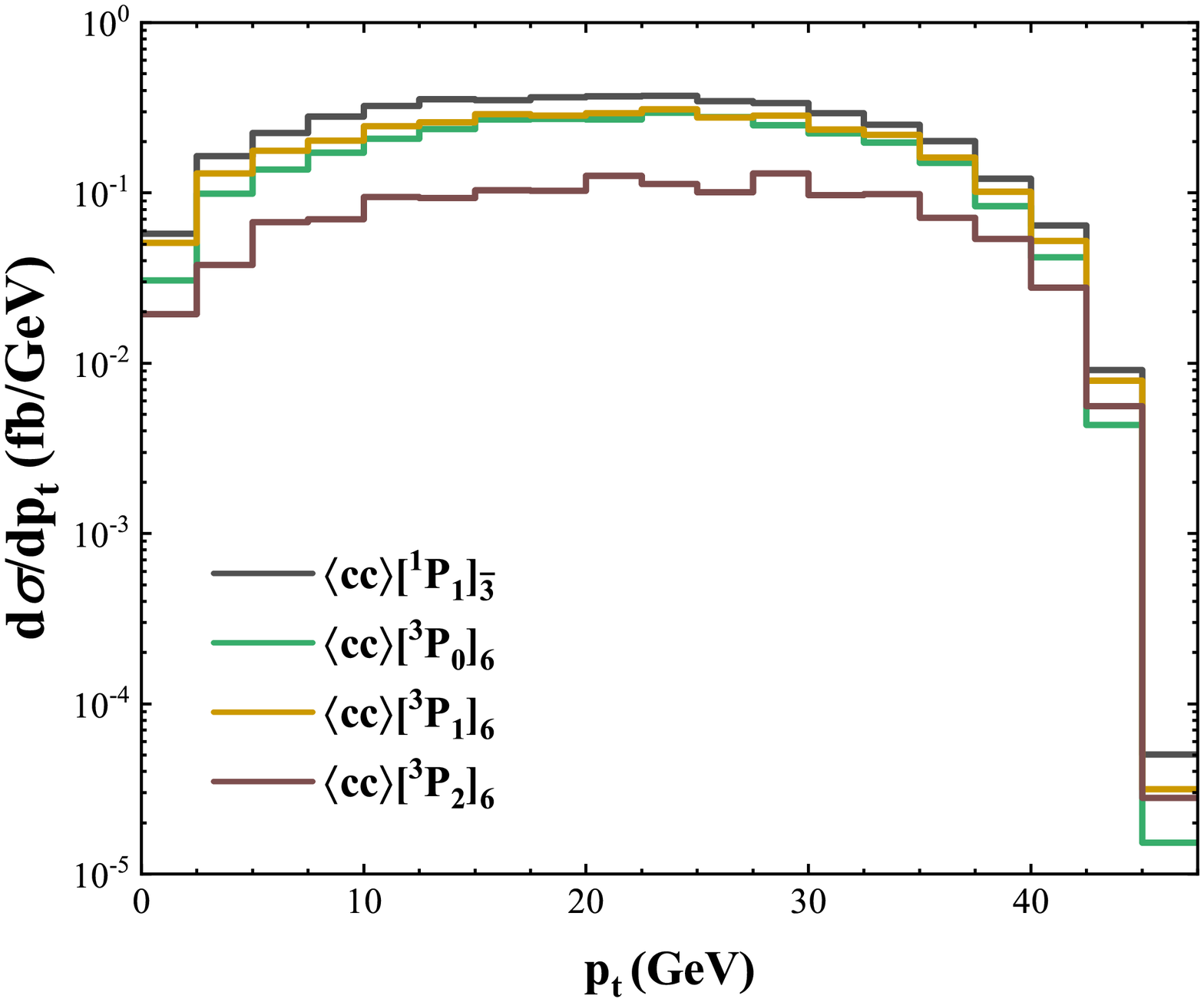}%
  \hspace{-0.7in}

  \hspace{-0.7in}
    \includegraphics[width=2.8in]{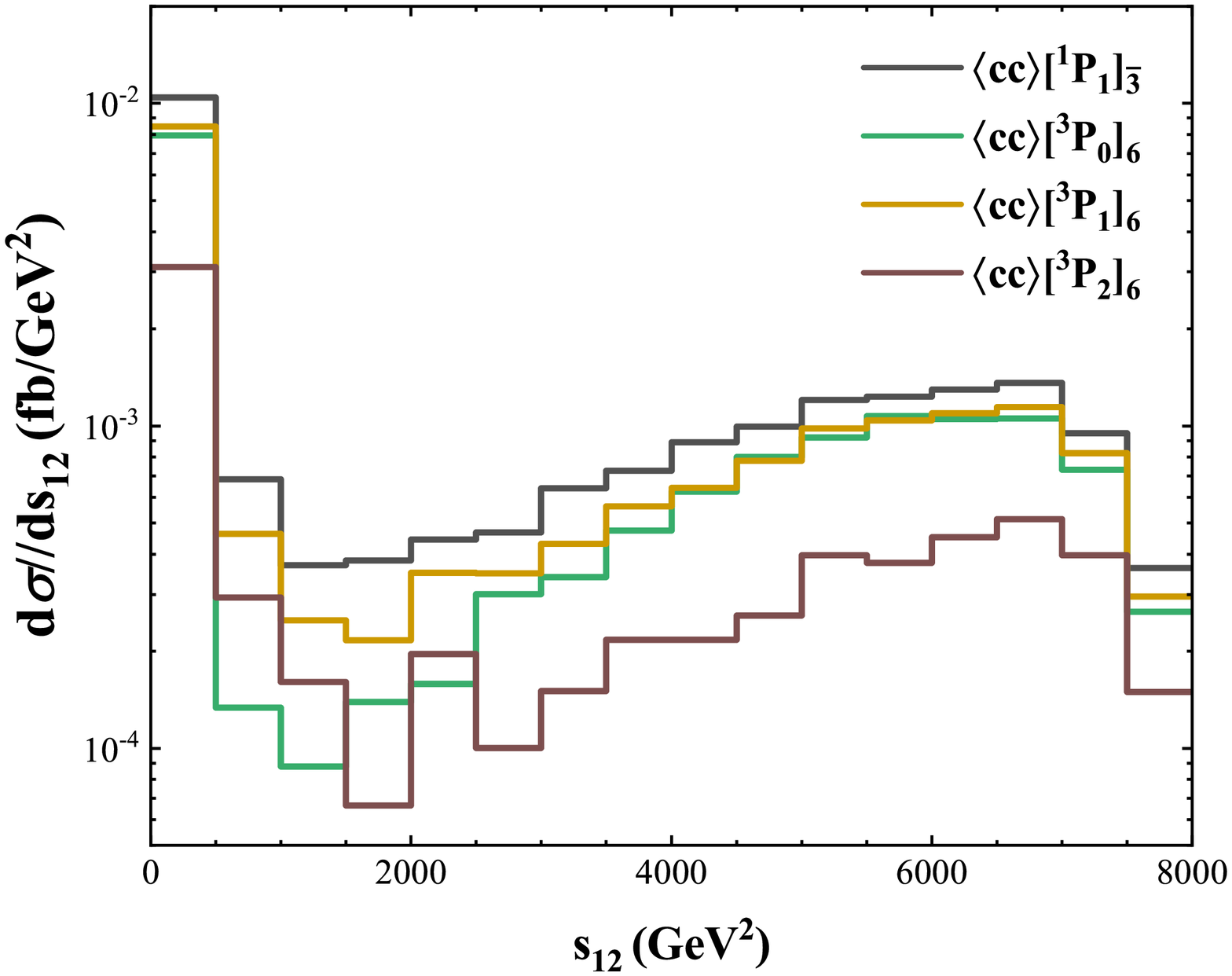}
  \hspace{-0.7in}
    \includegraphics[width=2.8in]{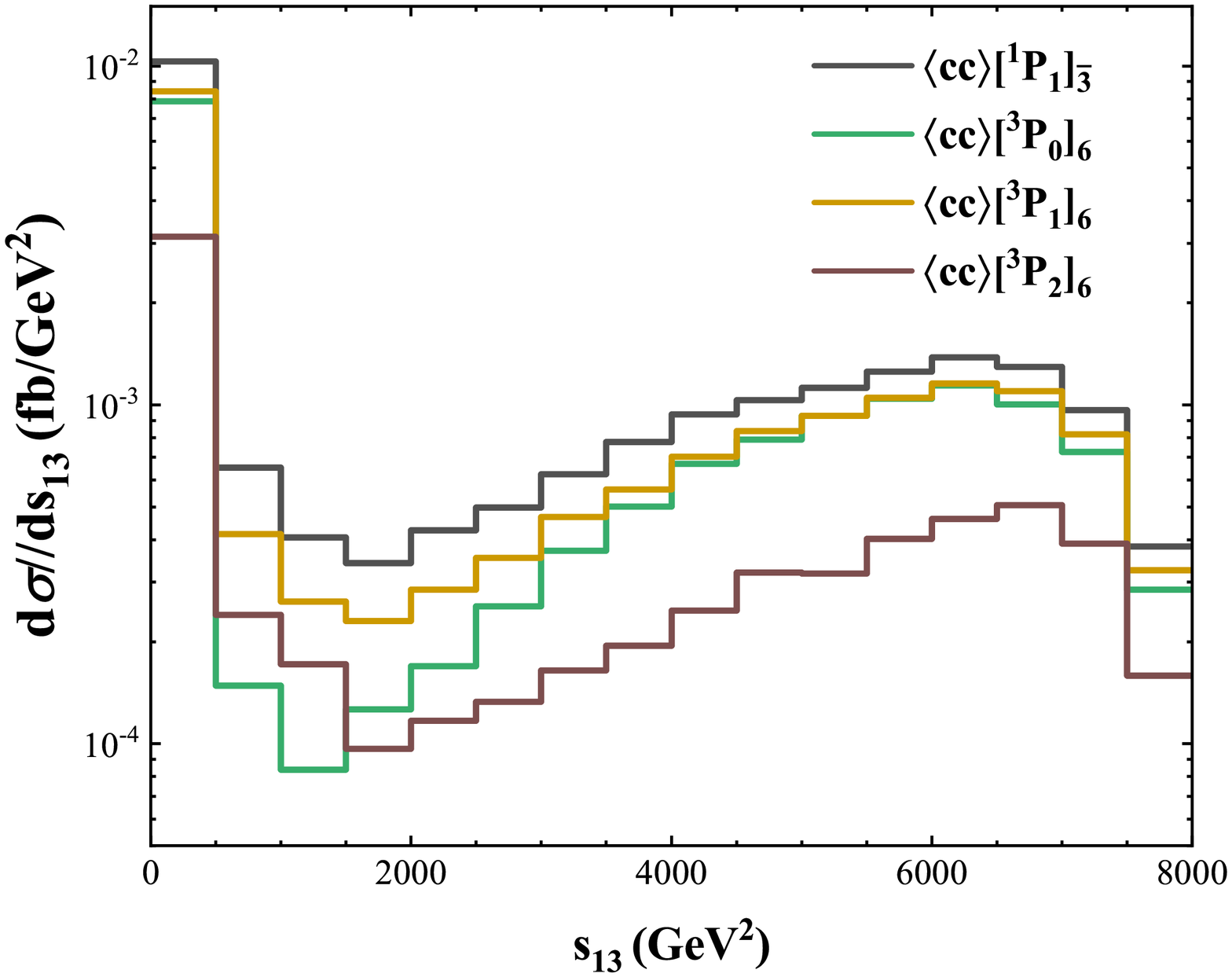}
  \hspace{-0.7in}
    \includegraphics[width=2.8in]{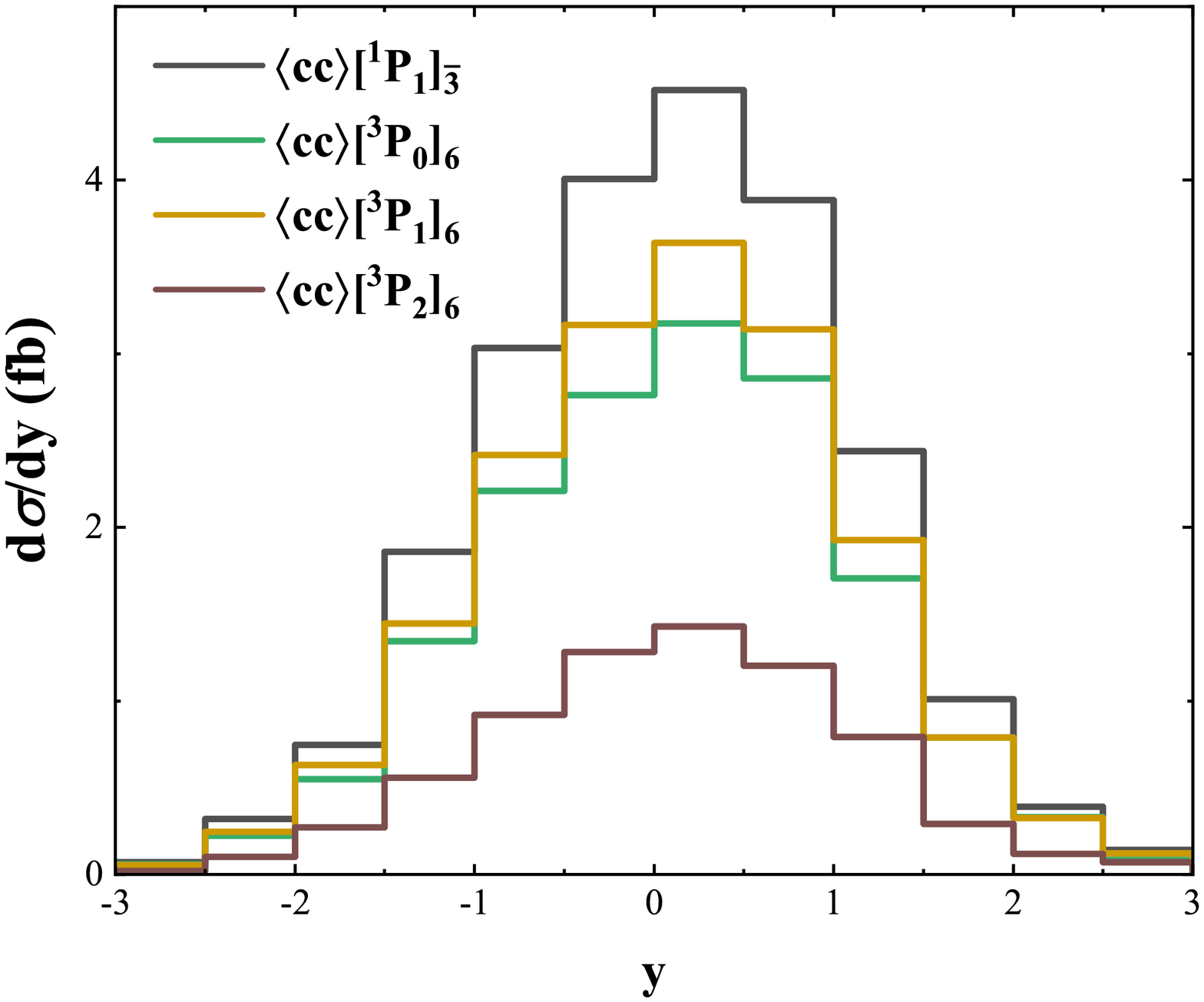}%
  \hspace{-0.7in}
  \caption{The  $\rm cos \rm{\theta}_{12}$,  $\rm cos \theta_{13}$, $\rm p_t$, $\rm s_{12}$,  $\rm s_{13}$, and y distributions for the $P$-wave $\Xi_{cc}$ production with intermediate $\langle cc\rangle[n]$ diquark at the Super-$Z$ factory.}
  \label{Xicc}
\end{figure}

\begin{figure}[htb]
  \centering
  \hspace{-0.7in}
    \includegraphics[width=2.8in]{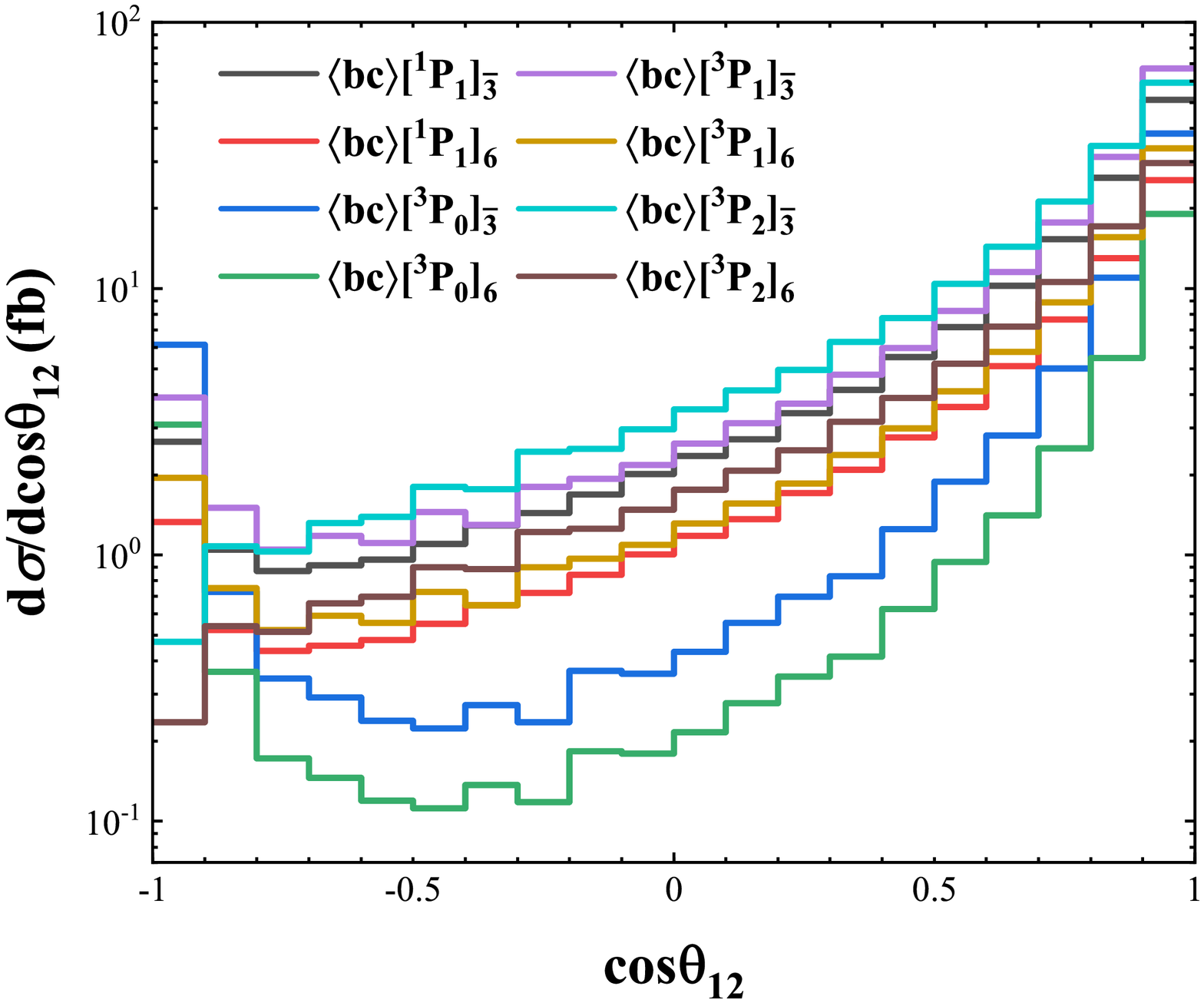}
  \hspace{-0.7in}
    \includegraphics[width=2.8in]{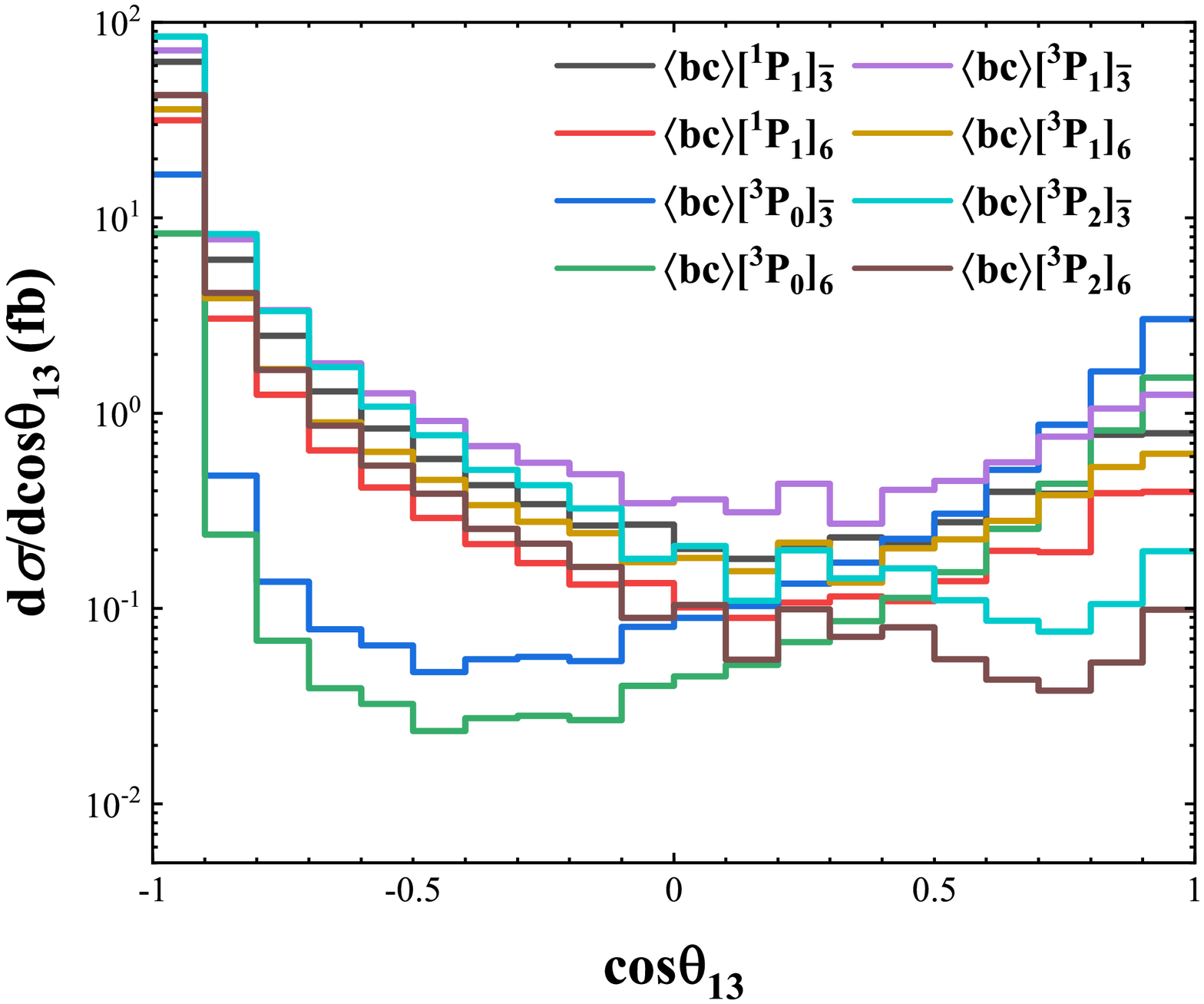}
  \hspace{-0.7in}
    \includegraphics[width=2.8in]{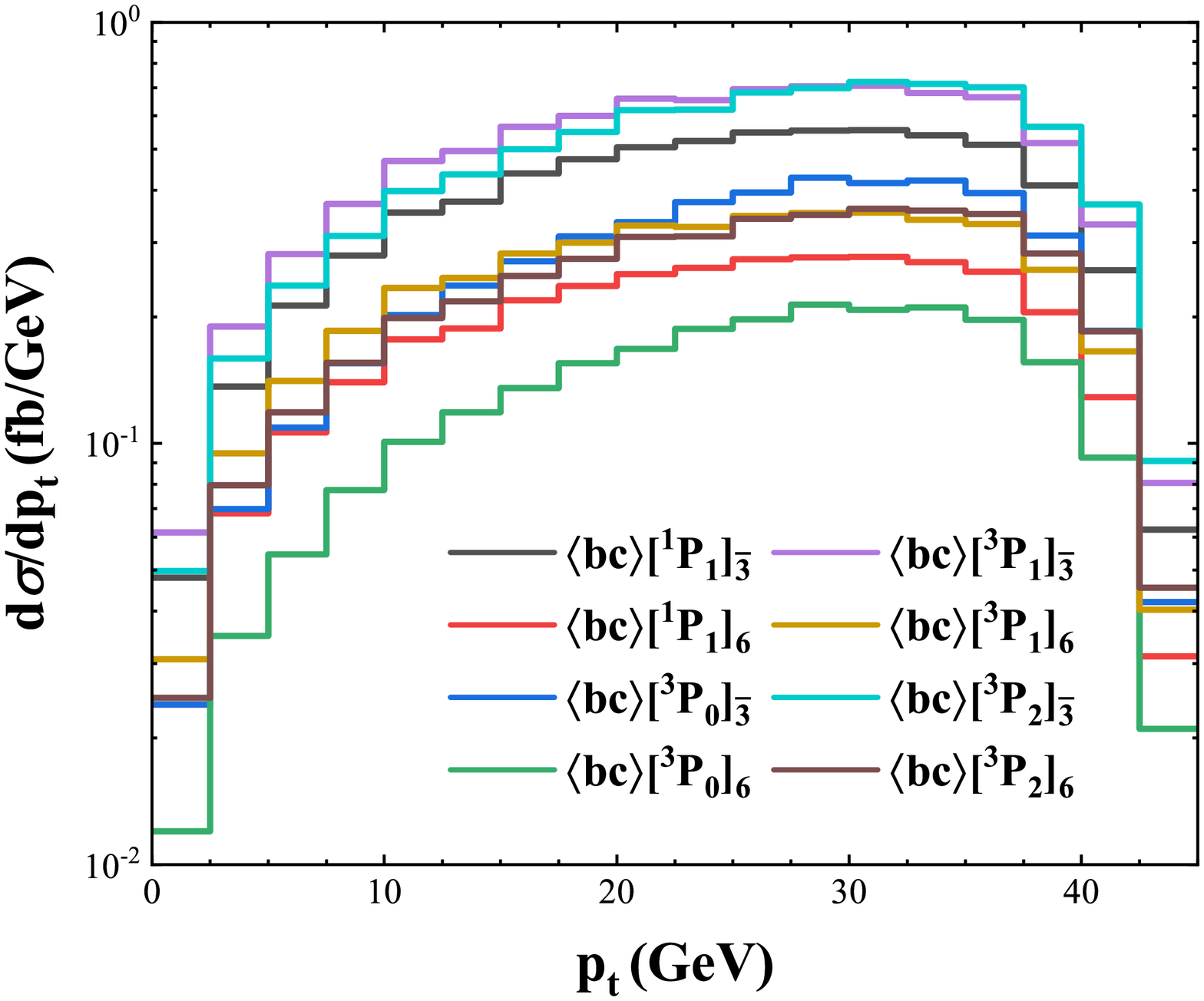}%
  \hspace{-0.7in}

  \hspace{-0.7in}
    \includegraphics[width=2.78in]{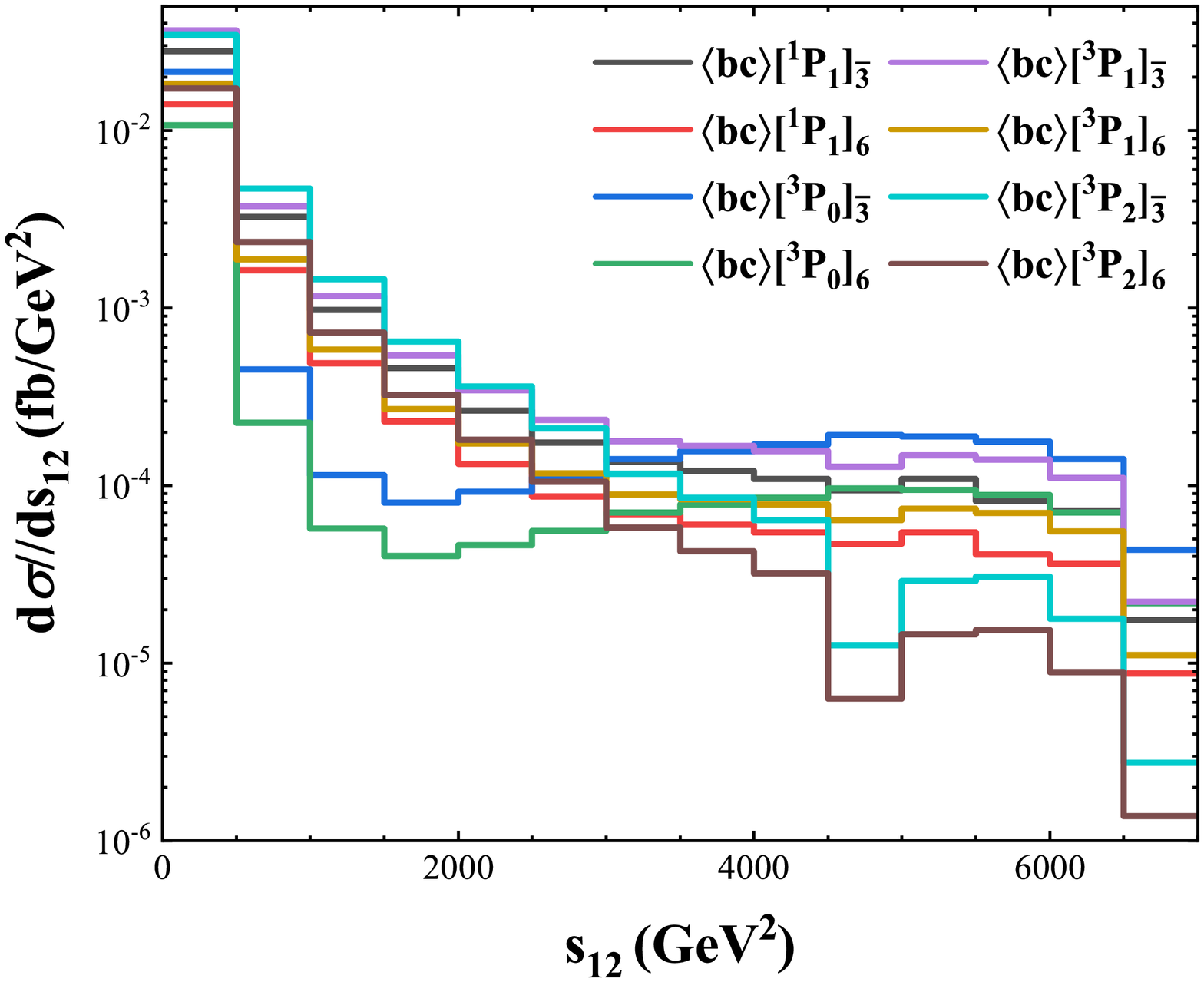}
  \hspace{-0.7in}
    \includegraphics[width=2.78in]{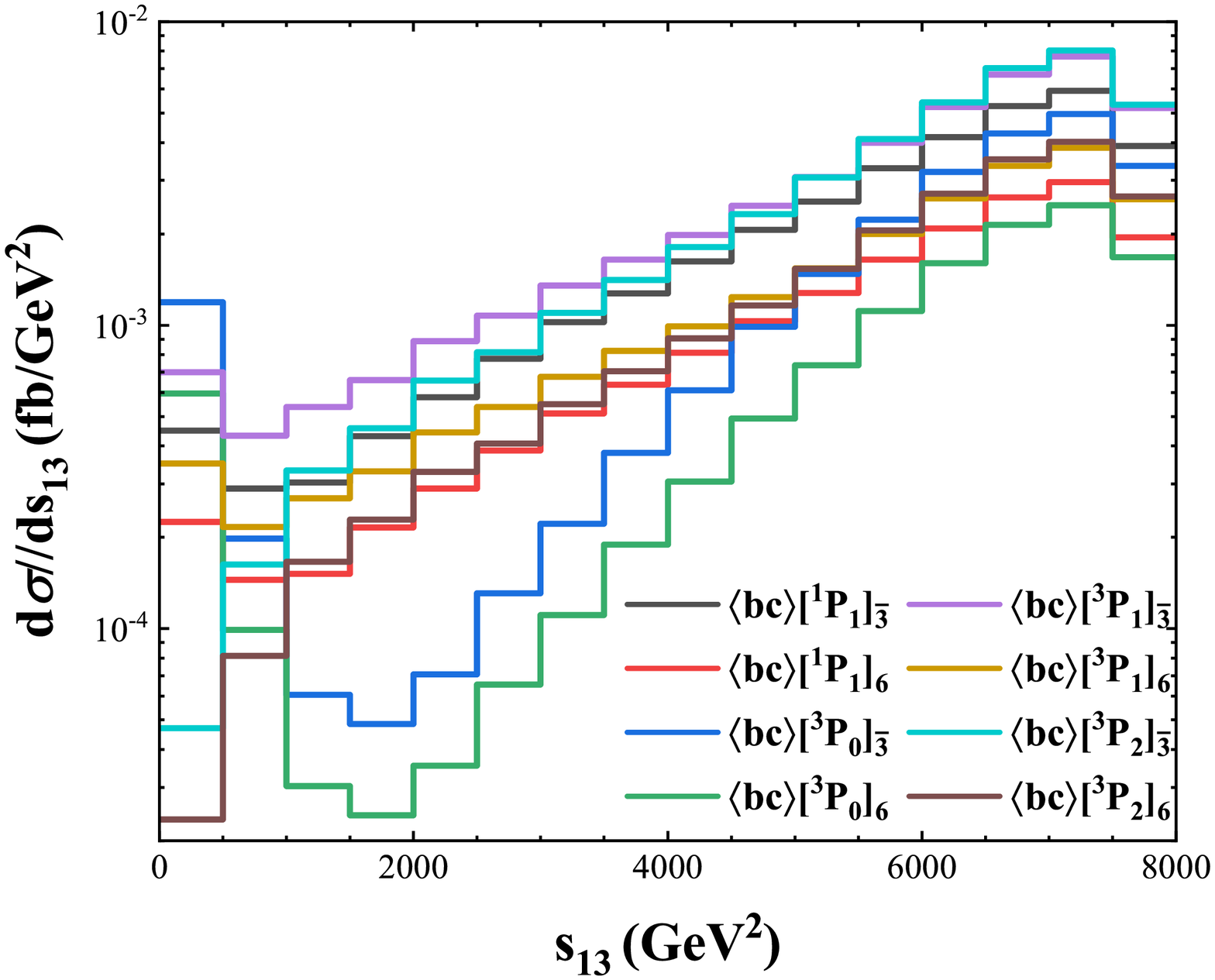}
  \hspace{-0.7in}
    \includegraphics[width=2.78in]{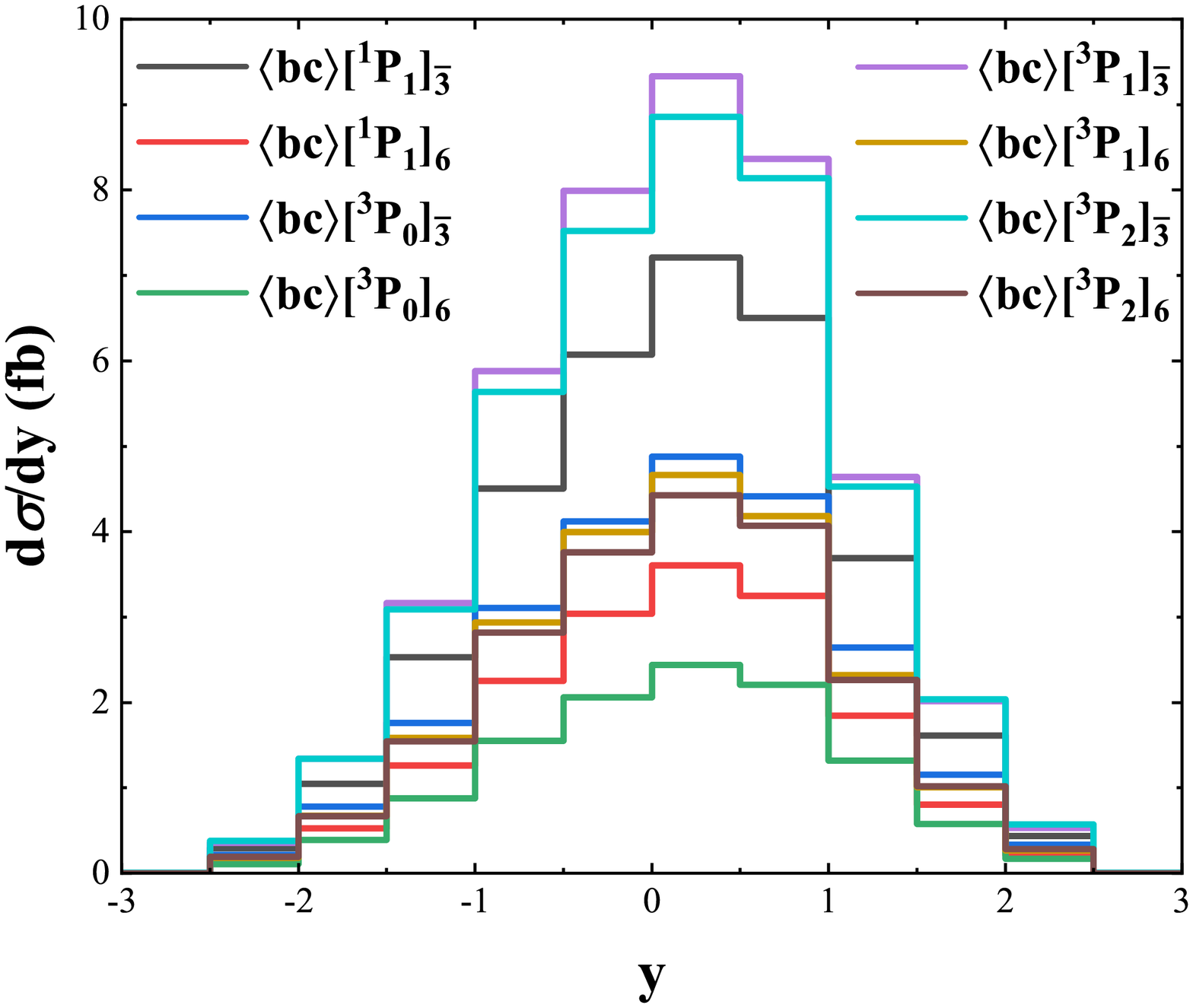}%
  \hspace{-0.7in}
  \caption{The $\rm cos \rm{\theta}_{12}$,  $\rm cos \theta_{13}$, $\rm p_t$, $\rm s_{12}$,  $\rm s_{13}$,  and y distributions for the $P$-wave $\Xi_{bc}$ production with intermediate $\langle bc\rangle[n]$ diquark at the Super-$Z$ factory.}
  \label{Xibc}
\end{figure}

\begin{figure}[htb]
  \centering
  \hspace{-0.7in}
    \includegraphics[width=2.8in]{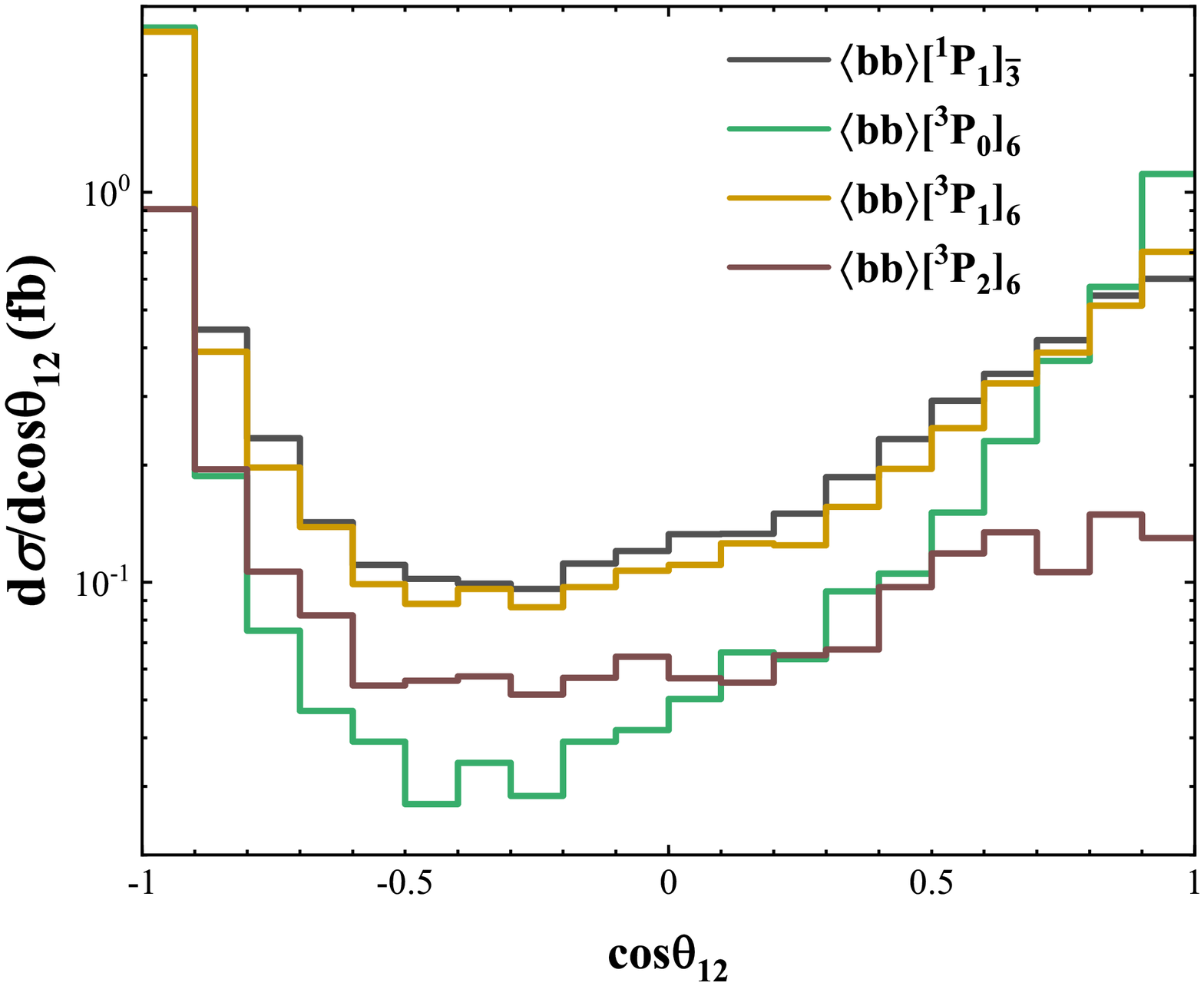}
  \hspace{-0.7in}
    \includegraphics[width=2.8in]{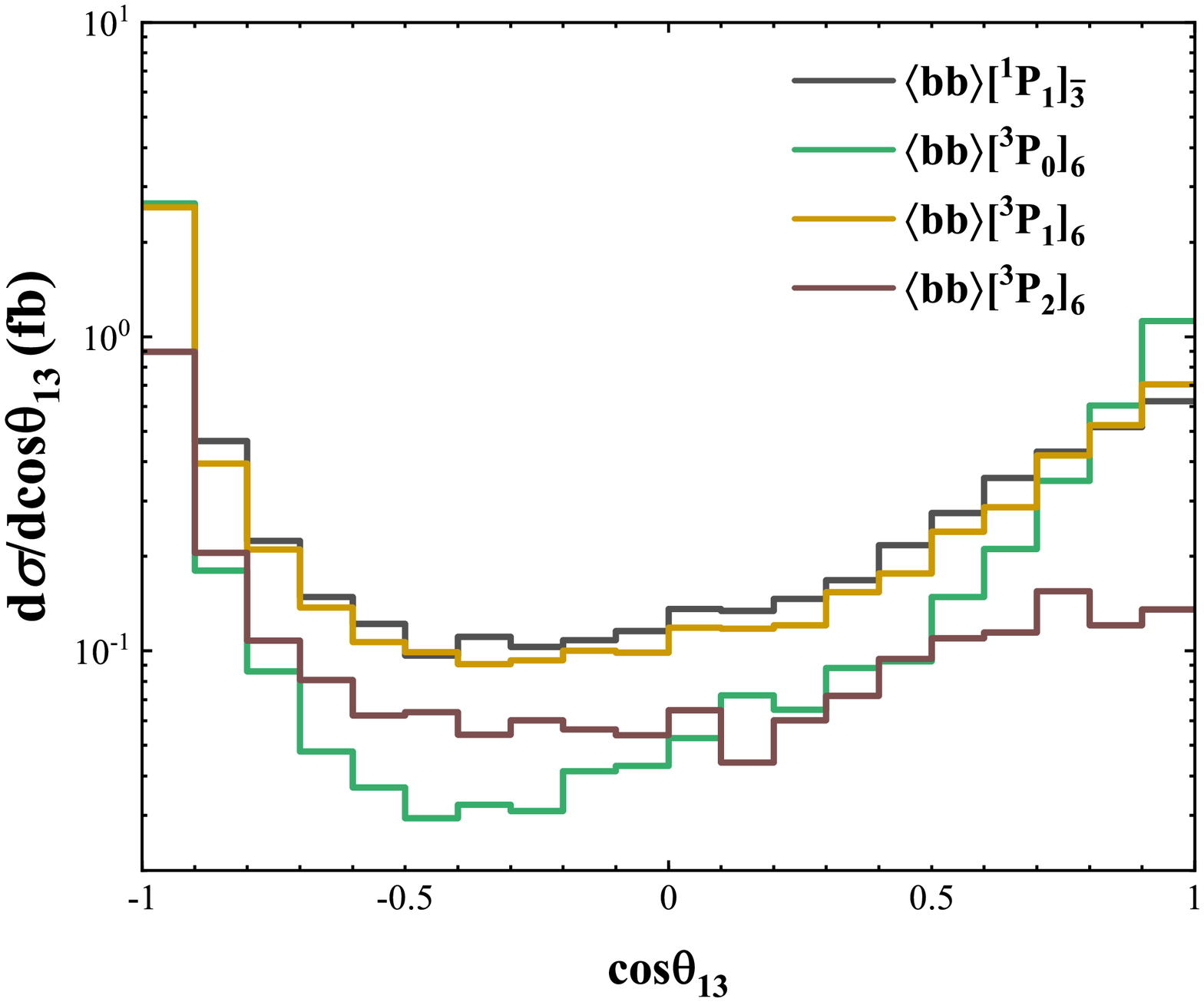}
  \hspace{-0.7in}
    \includegraphics[width=2.8in]{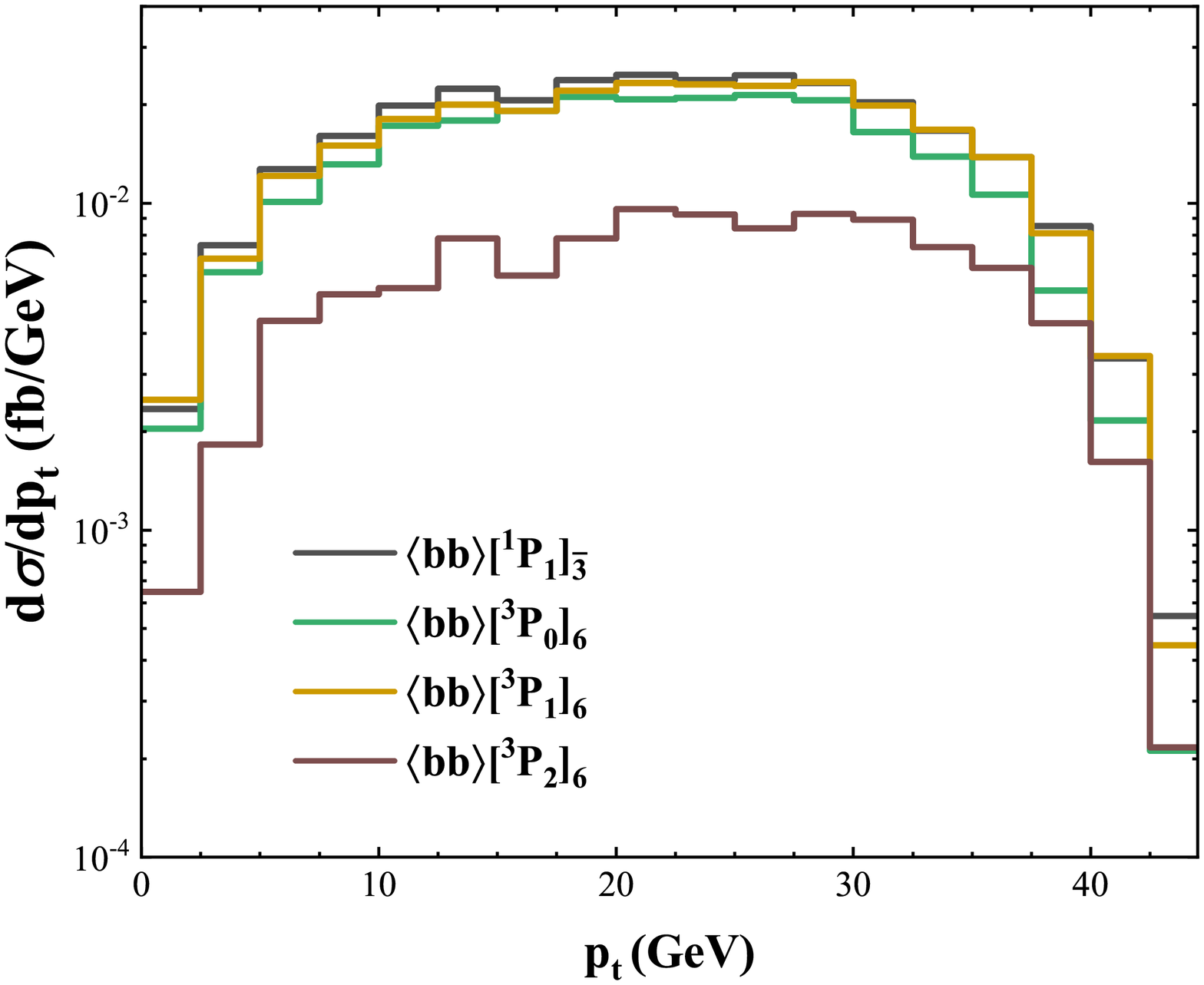}%
  \hspace{-0.7in}

  \hspace{-0.7in}
    \includegraphics[width=2.8in]{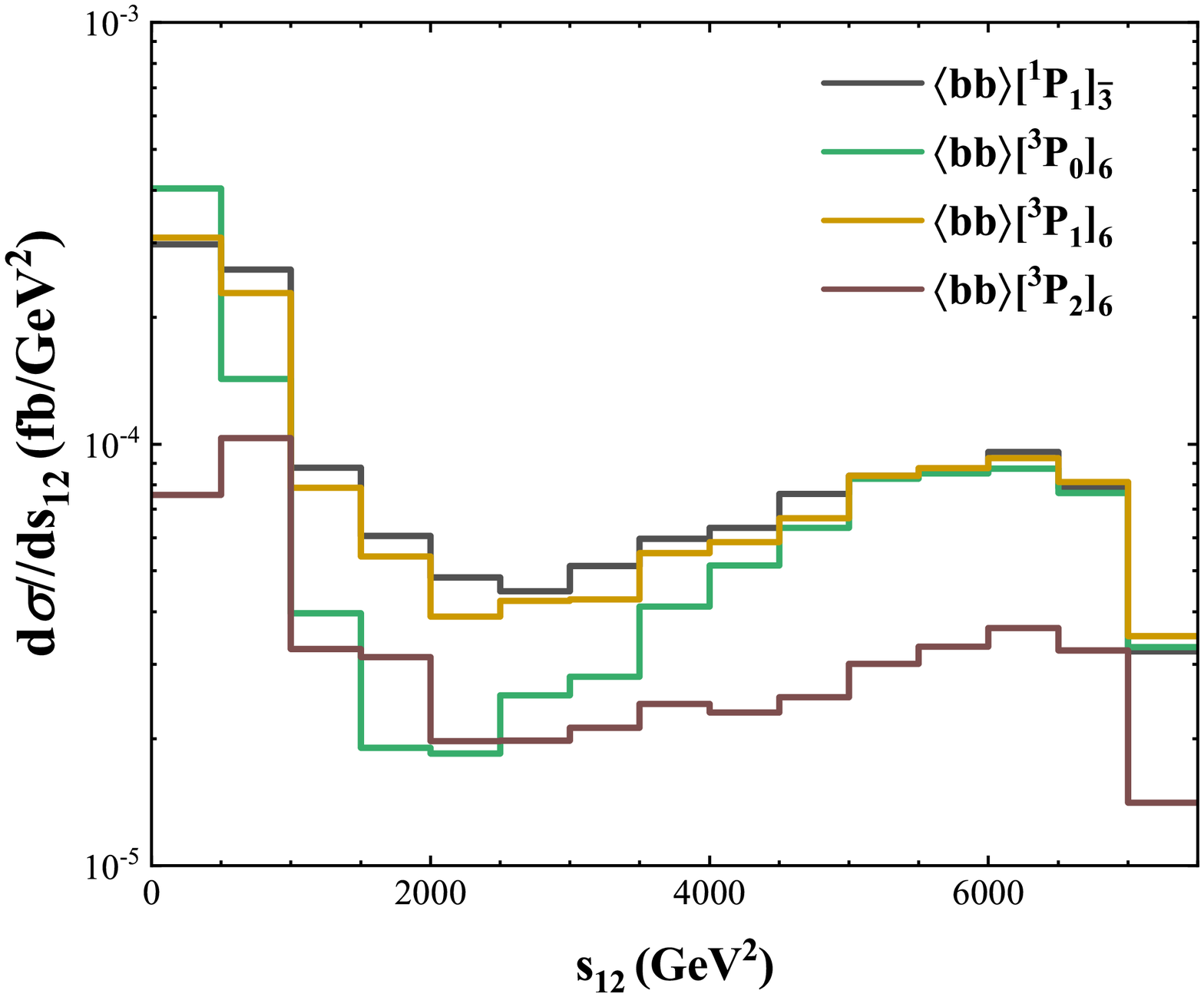}
  \hspace{-0.7in}
    \includegraphics[width=2.8in]{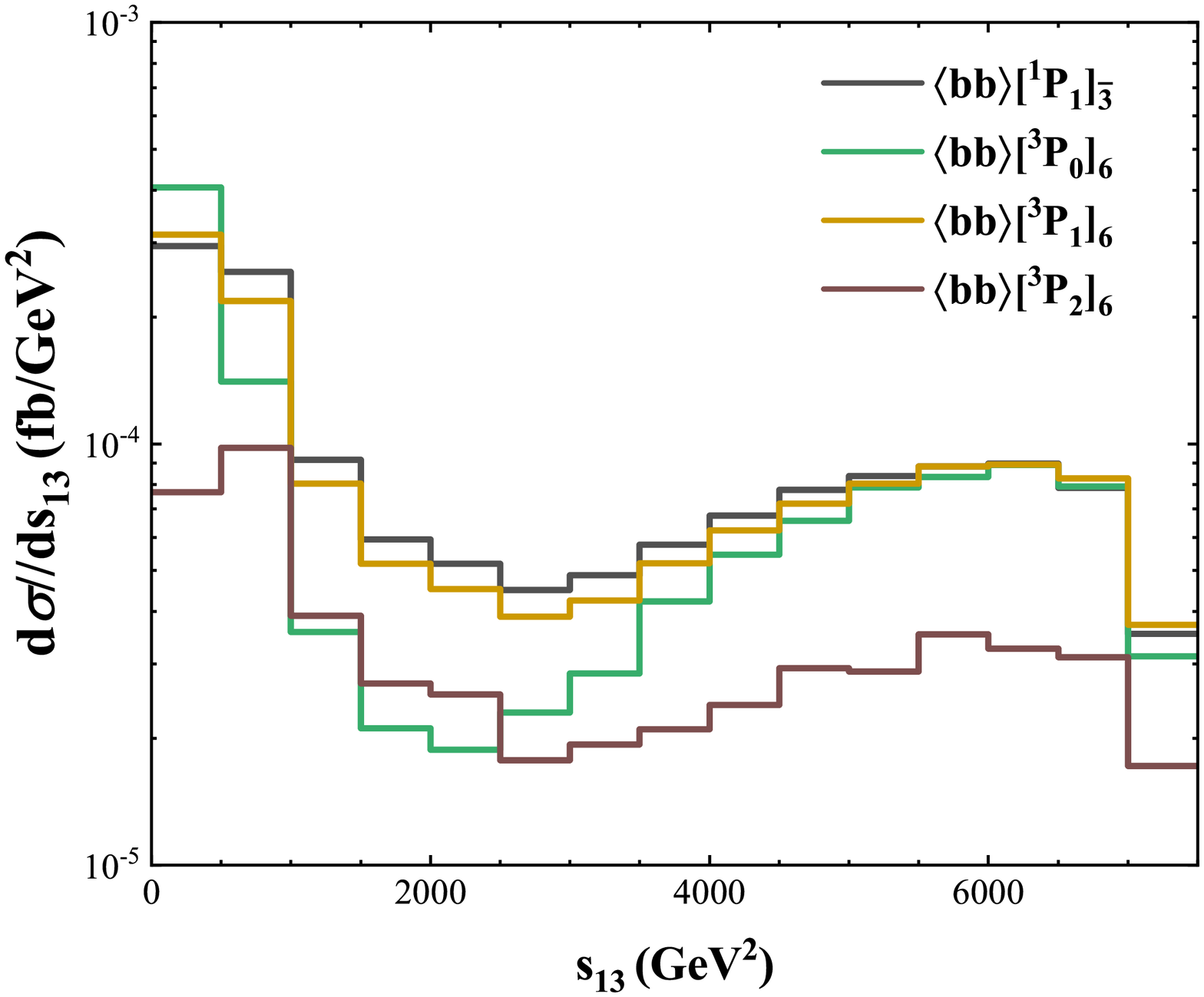}
  \hspace{-0.7in}
    \includegraphics[width=2.8in]{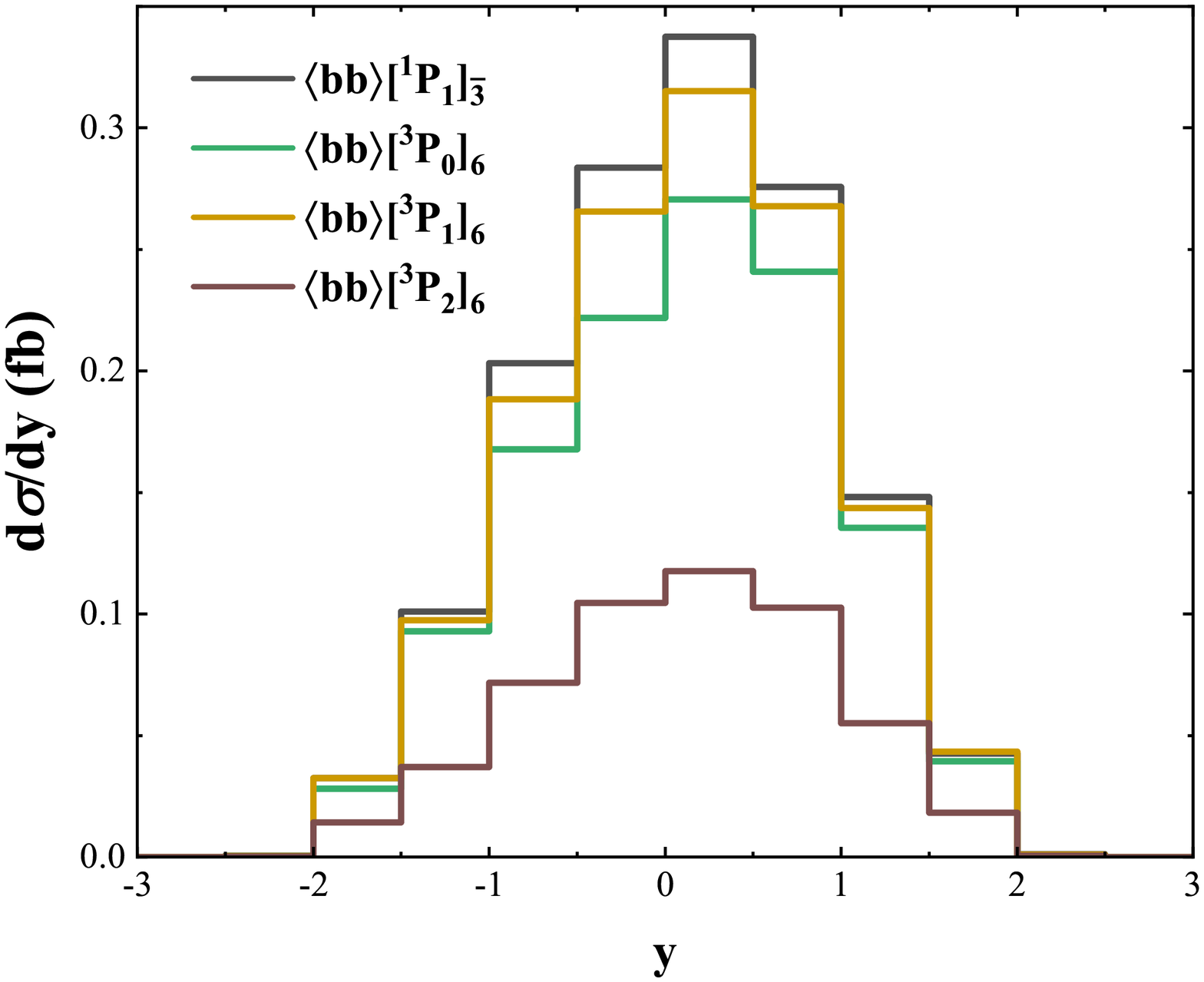}%
  \hspace{-0.7in}
  \caption{The $\rm cos \rm{\theta}_{12}$, $\rm cos \theta_{13}$, $\rm p_t$, $\rm s_{12}$,  $\rm s_{13}$,  and y distributions for the $P$-wave $\Xi_{bb}$ production with intermediate $\langle bb\rangle[n]$ diquark at the Super-$Z$ factory.}
  \label{Xibb}
\end{figure}

A large produced events of the doubly heavy baryons enable us to further study the kinematic properties of $\Xi_{QQ^{\prime}}$, such as the transverse momentum ($ \rm p_t$), rapidity (y), angular and invariant mass distributions, which are presented in Figs.~(\ref{Xicc}-\ref{Xibb}) with different intermediate $\langle cc\rangle[n]$, $\langle bc\rangle[n]$, and $\langle bb\rangle[n]$ diquarks, respectively. The definition of the invariant mass is s$_{ij}=(k_i+k_j)^2$, and $\theta_{ij}$ is the angle between momentum $\vec {k_i}$ and $\vec {k_j}$.

Figs.~(\ref{Xicc}-\ref{Xibb}) indicate that the distributions for the considered excited $\Xi_{cc}$ and $\Xi_{bb}$ with different $P$-wave configurations of diquark have similar behaviors, while they are slightly different from $\Xi_{bc}$ production.
d$\sigma/\rm {dcos} \theta_{12}$ (d$\sigma/\rm {dcos} \theta_{13}$) could reach its maximum when excited $\Xi_{QQ^{\prime}}$ and $\bar {Q^{\prime}}$ ($\bar{Q}$) move in the same direction or are back to back. The same conclusion can be drawn from the invariant mass differential distributions d$\sigma/\rm {s}_{12}$ and d$\sigma/\rm {s}_{13}$. The transverse momentum distribution, d$\sigma/\rm {dp} _{t}$, shall increase and then decreases with increasing p$_t$, i.e., the contribution of d$\sigma/\rm {dp} _{t}$ is relatively insignificant in the small p$_{t}$ regions (p$_{t}<5~\rm GeV$) and large p$_{t}$ regions (p$_{t}>40~\rm GeV$). From the figures of rapidity distributions, we can see that there are obvious backward-forward asymmetry, which is mainly caused by the exchange of $Z^0$ boson in the production of doubly heavy baryons at the Super-$Z$ factory. And the small rapidity regions ($\left| y\right|<2.0$) make the major contribution to d$\sigma/\rm {dy}$.

Then the theoretical uncertainty from the mass of heavy quark ($m_c$ and $m_b$) would be given in Tables \ref{unmc} and \ref{unmb} by varying $m_c= 1.8\pm 0.3$ GeV and $m_b= 5.1\pm 0.4$ GeV, separately, while other parameters remain at their central values. The cross sections for $\Xi_{QQ^{\prime}}$ production from different diquark configurations are all listed when $m_c=1.5$, $1.65$, $1.8$, $1.95$, and $2.1$ GeV and $m_b=4.7$, $4.9$, $5.1$, $5.3$, and $5.5$ GeV. 
According to Tables \ref{unmc} and \ref{unmb}, it is evident that all the cross sections decrease
with the increment of $m_c$. Specifically, compared with the result of the central value $\sigma_{m_c=1.8~\rm GeV}$ for $\Xi_{cc}$ production, the result of $m_c=1.5$, $1.65$, $1.95$, and $2.1$ GeV increases by 151.35$\%$, 55.31$\%$, -33.35$\%$, and -54.25$\%$, respectively. While for the production of $\Xi_{bc}$, the result of $m_c=1.5$, $1.65$, $1.95$, and $2.1$ GeV changes by 140.56$\%$, 51.70$\%$, -31.57$\%$, and -51.65$\%$, respectively. 
Whereas, the change of $m_b$ has a little effect on $\sigma_{\Xi_{bc}}$, and as compared to the result of the central value $\sigma_{m_b=5.1~\rm GeV}$, the result of $m_b=4.7$, $4.9$, $5.3$, and $5.5$ GeV increases by 3.54$\%$, 1.66$\%$, -1.49$\%$, and -2.82$\%$, respectively. All the cross sections for $\Xi_{bb}$ production decrease with the increment of $m_b$. Explicitly, the $\sigma$ of $m_b=4.7$, $4.9$, $5.3$, and $5.5$ GeV changes by 54.82$\%$, 23.93$\%$, -18.72$\%$, and -33.49$\%$, respectively, compared with $\sigma_{m_b=5.1~\rm GeV}$.

\begin{table}[htb]
\caption{Uncertainty of cross sections (in unit: fb) by varying $m_c= 1.8\pm0.3$ GeV. }
\centering
\renewcommand\arraystretch{1.2}
\begin{tabular}{|c|c|c|c|c|c|} 
\hline
$m_c$~(GeV)                                          & 1.5    & 1.65    & 1.8    & 1.95    & 2.1    \\ 
\hline\hline
$\sigma_{\Xi_{cc}}([^1P_1]_{\overline{\mathbf{3}}})$ & 28.86  & 17.80  & 11.43  & 7.60  & 5.20   \\
\hline
$\sigma_{\Xi_{cc}}([^3P_0]_{\mathbf{6}})$            & 20.57  & 12.74  & 8.23  & 5.50  & 3.78   \\
\hline
$\sigma_{\Xi_{cc}}([^3P_1]_{\mathbf{6}})$            & 23.00  & 14.20  & 9.14  & 6.09  & 4.18   \\
\hline
$\sigma_{\Xi_{cc}}([^3P_2]_{\mathbf{6}})$            & 8.95  & 5.54  & 3.58  & 2.39  & 1.65   \\
\hline\hline
$\sigma_{\Xi_{bc}}([^1P_1]_{\overline{\mathbf{3}}})$ & 42.05  & 26.37  & 17.28  & 11.75  & 8.25   \\ 
\hline
$\sigma_{\Xi_{bc}}([^1P_1]_{\mathbf{6}})$            & 21.02  & 13.18  & 8.64  & 5.87  & 4.13   \\
\hline
$\sigma_{\Xi_{bc}}([^3P_0]_{\overline{\mathbf{3}}})$ & 24.59  & 16.82  & 11.95  & 8.77  & 6.60   \\ 
\hline
$\sigma_{\Xi_{bc}}([^3P_0]_{\mathbf{6}})$            & 12.29  & 8.41  & 5.98  & 4.38  & 3.30   \\
\hline
$\sigma_{\Xi_{bc}}([^3P_1]_{\overline{\mathbf{3}}})$ & 51.40  & 33.11  & 22.24  & 15.48  & 11.11   \\
\hline
$\sigma_{\Xi_{bc}}([^3P_1]_{\mathbf{6}})$            & 25.70  & 16.56  & 11.12  & 7.74  & 5.55   \\
\hline
$\sigma_{\Xi_{bc}}([^3P_2]_{\overline{\mathbf{3}}})$ & 57.34  & 34.30  & 21.43  & 13.89  & 9.29   \\
\hline
$\sigma_{\Xi_{bc}}([^3P_2]_{\mathbf{6}})$            & 28.67  & 17.15  & 10.72  & 6.95  & 4.65   \\
\hline
\end{tabular}
\label{unmc}
\end{table}

\begin{table}[htb]
\caption{Uncertainty of cross sections (in unit: fb) by varying $m_b= 5.1\pm 0.4$ GeV. }
\centering
\renewcommand\arraystretch{1.2}
\begin{tabular}{|c|c|c|c|c|c|} 
\hline
$m_b$(GeV)                                           & 4.7    & 4.9    & 5.1    & 5.3    & 5.5     \\ 
\hline\hline
$\sigma_{\Xi_{bc}}([^1P_1]_{\overline{\mathbf{3}}})$ & 17.80  & 17.52  & 17.28  & 17.05  & 16.85   \\ 
\hline
$\sigma_{\Xi_{bc}}([^1P_1]_{\mathbf{6}})$            & 8.90  & 8.76  & 8.64  & 8.53  & 8.43   \\ 
\hline
$\sigma_{\Xi_{bc}}([^3P_0]_{\overline{\mathbf{3}}})$ & 13.21  & 12.55  & 11.95  & 11.42  & 10.94   \\ 
\hline
$\sigma_{\Xi_{bc}}([^3P_0]_{\mathbf{6}})$            & 6.60  & 6.27  & 5.98  & 5.71  & 5.47   \\
\hline
$\sigma_{\Xi_{bc}}([^3P_1]_{\overline{\mathbf{3}}})$ & 23.43  & 22.81  & 22.24  & 21.73  & 21.27   \\
\hline
$\sigma_{\Xi_{bc}}([^3P_1]_{\mathbf{6}})$            & 11.71  & 11.40  & 11.12  & 10.87  & 10.63   \\
\hline
$\sigma_{\Xi_{bc}}([^3P_2]_{\overline{\mathbf{3}}})$ & 21.05  & 21.24  & 21.43  & 21.61  & 21.79   \\
\hline
$\sigma_{\Xi_{bc}}([^3P_2]_{\mathbf{6}})$            & 10.52  & 10.62  & 10.72  & 10.81  & 10.89   \\
\hline\hline
$\sigma_{\Xi_{bb}}([^1P_1]_{\overline{\mathbf{3}}})$ & 1.14  & 0.91  & 0.73  & 0.59  & 0.48   \\
\hline
$\sigma_{\Xi_{bb}}([^3P_0]_{\mathbf{6}})$            & 0.94  & 0.76  & 0.61  & 0.50  & 0.41   \\
\hline
$\sigma_{\Xi_{bb}}([^3P_1]_{\mathbf{6}})$            & 1.06  & 0.85  & 0.69  & 0.56  & 0.46   \\
\hline
$\sigma_{\Xi_{bb}}([^3P_2]_{\mathbf{6}})$            & 0.41  & 0.33  & 0.27  & 0.22  & 0.18   \\
\hline
\end{tabular}
\label{unmb}
\end{table}

Then, the theoretical uncertainty of the non-perturbative matrix element $\langle{\cal O}^H[n]\rangle$ would be discussed. For the diquark in color-antitriplet $\bar{\mathbf 3}$, it can be approximately obtained from the original of the radial wave function for $S$-wave and its first derivative for $P$-wave. At present, there are two common views regarding the transition probability of the color-sextuplet $\mathbf 6$ \cite{Ma:2003zk}. A conventional view claims that the contribution of color-sextuplet state is as important as that of color-antitriplet by the velocity scaling rule, resulting in the same transition probability of color-sextuplet and -antitriplet, $h_6\simeq h_{\bar 3}$ \cite{Jiang:2012jt,Niu:2018ycb}. Another view suggests that the contribution of color-sextuplet state should be suppressed to the color-antitriplet by order $v^2$, that is, $h_6\simeq v^2 h_{\bar 3}$, or it could even be ignored \cite{Niu:2019xuq,Zheng:2015ixa}. Based on this situation, the produced events of doubly heavy baryons from intermediate diquark in color-sextuplet would also be suppressed, and our numerical results can be treated as an upper limit at the Super-$Z$ factory. 

Finally, we present a quantitive uncertainty analysis of the radial wave functions at the origin and their first derivatives for doubly heavy diquarks. At present, the first derivative of the radial wave function at the origin for the $\langle cc \rangle$-, $\langle bc \rangle$-, $\langle bb \rangle$-diquark systems have been calculated under two different potential models, i.e., $\rm K^2$O \cite{Kiselev:2002iy} potential and Buchm$\rm \ddot{u}$ller-Tye ($\rm B.T.$) \cite{Kiselev:2001fw} potential, and the results are listed in Table~\ref{radial}. We can see that the results obtained by the power-low potential are larger than those calculated under $\rm K^2$O and B.T. potential models. The radial wave functions calculated with the $\rm K^2$O potential and B.T. potential are very similar.
To obtain the uncertainty of the radial wave functions and their first derivative at the origin of the diquark systems, we take the average of the radial wave functions at the origin calculated with different potential models as the central value, and the uncertainties of the radial wave functions at the origin are $0.582^{+0.118}_{-0.059}$, $0.784^{+0.120}_{-0.062}$, and $1.358^{+0.024}_{-0.013}$ $\rm GeV^{3/2}$ for the $\langle cc \rangle$-, $\langle bc \rangle$-, and $\langle bb \rangle$-diquark systems respectively. Similarly for the first derivatives of the radial wave functions at the origin, the uncertainties are $0.115\pm0.013$, $0.201\pm0.01$, and $0.479\pm 0$ $\rm GeV^{5/2}$ for the $\langle cc \rangle$-, $\langle bc \rangle$-, and $\langle bb \rangle$-diquark systems respectively. The uncertainties of the total cross sections caused by the radial wave functions at the origin are $604.93^{+129.97}_{-117.3}$, $1728.90^{+533.81}_{-458.14}$ and $39.81^{+1.36}_{-1.33}$ fb for $\Xi_{cc}$, $\Xi_{bc}$, and $\Xi_{bb}$ production, respectively.
However, the calculation of the radial wave function is non-trivial, and there are some naive assumptions for estimating its value at the origin \cite{Falk:1993gb,Baranov:1995rc}, which might introduce greater uncertainty. Fortunately, these radial wave functions are global factors in our calculations, and it is easy to update the numerical results in the future if more accurate radial wave function values are available.

\begin{table}[htb]
\caption{Radial wave functions at the origin and their first derivatives of the $\langle cc \rangle$-, $\langle bc \rangle$-, and $\langle bb \rangle$-diquark systems. }
\centering
\begin{tabular}{|c||c|c|c||c|c|c|}
\hline
\multirow{2}*{State} & \multicolumn{3}{|c||}{S-wave} & \multicolumn{3}{c|}{P-wave} \\
\cline{2-7}
~& $\langle cc \rangle$ &$\langle bc \rangle$ &$\langle bb \rangle$ & $\langle cc \rangle$ &$\langle bc \rangle$ &$\langle bb \rangle$  \\
\hline\hline
power-low \cite{Baranov:1995rc} & 0.700 & 0.904 & 1.382  & -  & - & - \\
\hline
$\rm K^2$O \cite{Kiselev:2002iy} & 0.523 & 0.722 & 1.345  & 0.102  & 0.200  & 0.479  \\
\hline
B.T.\cite{Kiselev:2001fw}  & 0.530 & 0.726 & 1.346  & 0.128  & 0.202  & 0.479 \\
\hline
\end{tabular}
\label{radial}
\end{table}

\section{summary}\label{summary}

In this manuscript, the production of excited doubly heavy baryon $\Xi_{QQ^{\prime}}$ with intermediate diquark in $P$-wave has been analyzed at the Super-$Z$ factory using NRQCD theory. At the collision energy $\sqrt s=m_Z$, the contribution of the intermediate $\gamma$ propagator is negligible compared to that of the $Z$ propagator through the annihilation process, $e^{+} + e^{-}\rightarrow \langle QQ^{\prime}\rangle[n] \rightarrow \Xi_{QQ^{\prime}} +\bar{Q} +\bar{Q^{\prime}}$, where $Q^{(\prime)}$ = $b$ or $c$ quark.

The total cross sections and relevant events for the production of excited $\Xi_{cc}$, $\Xi_{bc}$, and $\Xi_{bb}$ baryons from the $P$-wave diquark state at the Super-$Z$ factory with ${\cal L} \simeq 10^{34}~{\rm cm}^{-2} {\rm s}^{-1}$ in one operational year are
\begin{eqnarray}
\sigma _{\Xi_{cc}}&=&32.38~\rm fb,~~~~~~N_{\Xi_{cc}}=3.24 \times10^{3};\nonumber\\
\sigma _{\Xi_{bc}}&=&109.36~\rm fb, ~~~~N_{\Xi_{bc}}=1.09 \times10^{4};\nonumber\\
\sigma _{\Xi_{bb}}&=&2.30~\rm fb, ~~~~~~~N_{\Xi_{bb}}=2.30 \times10^{2};\nonumber
\end{eqnarray}
which are 3.97$\%$, 5.08$\%$, and 5.89$\%$ of the contribution of all summed $S$-wave, correspondingly. The produced events per year at the GigaZ are 0.7 times of that at the Super-$Z$ factory.
Assuming that all the considered excited states can decay into the ground state 100 $\%$, the total cross sections for $\Xi_{cc}$, $\Xi_{bc}$, and $\Xi_{bb}$ production are $848.03$ fb, $2260.51$ fb, and $41.17$ fb, respectively, which result in a large number of produced events up to $8.48 \times10^{4}$, $2.26 \times10^{5}$, and $4.12 \times10^{3}$, respectively. When the luminosity of the Super-$Z$ factory increases to $10^{36}~{\rm cm}^{-2} {\rm s}^{-1}$, the number of produced $\Xi_{QQ^{\prime}}$ events will increase by 100 times, showing that
the Super-$Z$ factory could be a potential platform for the study of doubly heavy baryons, and we expect more signals of doubly heavy baryons to be detected by the experiment.

To provide some guidance for experimental measurements, the relevant transverse momentum, rapidity, angular, and invariant mass distributions are presented. The largest contributions can be achieved when the excited $\Xi_{QQ^{\prime}}$ baryons and $\bar {Q^{\prime}}$ ($\bar{Q}$) are moving side by side or back to back. The contribution of d$\sigma/\rm {dp} _{t}$ is mainly presented in the region 5 GeV $<$ p$_{t}<40~\rm GeV$. There are obvious backward-forward asymmetry in the rapidity distributions, and the small rapidity ($\left| y\right|<2.0$) dominates the major contribution of d$\sigma/\rm {dy}$. 
 
Finally, we discussed the theoretical uncertainty from the heavy quark mass ($m_c$ and $m_b$), the non-perturbative matrix element, and the radial wave functions. The results indicate a monotonic decrease for $\sigma_{\Xi_{cc}}$ and $\sigma_{\Xi_{bc}}$ in all considered configurations with $m_c$. While the change of $m_b$ has little effect on $\sigma_{\Xi_{bc}}$. All the cross sections for $\Xi_{bb}$ production decrease with the increment of $m_b$. If the contribution of the color-sextuplet state should be suppressed to color-antitriplet by order $v^2$ or could even be ignored, the number of produced events of doubly heavy baryons from intermediate diquark in color-sextuplet would also be suppressed, and our numerical results can be treated as an upper limit at the Super-$Z$ factory. According to the current wave functions at the origin and their first derivatives calculated with different potential models, the uncertainties of the cross sections are still large for the production of $\Xi_{QQ^{\prime}}$.

\hspace{2cm}

{\bf Acknowledgements}: 
We would like to thank the fruitful discussion with Xu-Chang Zheng. This work was partially supported by the Natural Science Foundation of Guangxi (No. 2020GXNSFBA159003), the Guangxi Technology Base and Talent Subject (No. Guike AD20238014 and No. Guike AD23026182), and the National Natural Science Foundation of China (No.12005045). The research was also supported by the China Postdoctoral Science Foundation under Grant No. 2022TQ0012 and No. 2023M730097, and the Central Government Guidance Funds for Local Scientific and Technological Development, China (No. Guike ZY22096024).

\end{document}